\documentclass[Afour,sageh]{sagej}

\usepackage[utf8]{inputenc}
\usepackage{tikz}
\usetikzlibrary{positioning}
\usetikzlibrary{calc}
\usetikzlibrary{chains}
\usetikzlibrary{shapes}
\usepackage{booktabs}
\usepackage{graphics}
\usepackage{wrapfig}
\usepackage{placeins}
\usepackage{amsmath}
\usepackage{listings}
\usepackage{microtype}
\usepackage{hyperref}
\usepackage{hypernat}
\usepackage{ifthen}
\usepackage{amssymb}
\usepackage{algorithm}
\usepackage{algorithmic}
\usepackage[skins,listings]{tcolorbox}
\usepackage[normalem]{ulem}
\usepackage{float}
\usepackage{subcaption}
\usepackage[export]{adjustbox}
\usepackage{multirow}
\usepackage{varwidth}
\usepackage{url}

\usetikzlibrary{shapes.misc}
\tikzset{cross/.style={cross out, draw=black, minimum size=2*(#1-\pgflinewidth), inner sep=0pt, outer sep=0pt, thick}, cross/.default={0.16cm}}

\newfloat{algorithm}{t}{lop}

\def\loopy{\textsc{Loopy}}
\def\uipick{\textsc{UIPiCK}}
\def\perflex{\textsc{Perflex}}
\def\nvidiavolta{Nvidia Titan V}
\def\nvidiamaxwell{Nvidia GTX Titan X}
\def\nvidiakepler{Nvidia Tesla K40c}
\def\nvidiafermi{Nvidia Tesla C2070}
\def\amdradeon{AMD Radeon R9 Fury}

\definecolor{plotgreen}{RGB}{0,128,0}
\definecolor{plotblue}{RGB}{0,0,255}
\definecolor{plotred}{RGB}{255,0,0}
\definecolor{plotmagenta}{RGB}{191,0,191}

\newlength\q
\tcbset{listingbox/.style={%
    enhanced,
    boxsep=-9pt,
    left=12pt,
    arc=2pt,
    boxrule=0.8pt,
  }
}

\definecolor{green}{RGB}{0, 180, 0}

\lstdefinestyle{custompython}{%
    belowcaptionskip=1\baselineskip,
    breaklines=true,
    frame=none,
    xleftmargin=\parindent,
    language=Python,
    showstringspaces=false,
    basicstyle=\footnotesize\ttfamily,
    keywordstyle=\bfseries\color{green!40!black},
    commentstyle=\itshape\color{purple},
    identifierstyle=\color{black},
    stringstyle=\color{green!60!black},
    keywords=[2]{as,True,False},
    keywordstyle=[2]\bfseries\color{green!40!black},
    keywords=[3]{np,sp,plt},
    keywordstyle=[3]\bfseries\color{blue},
    numbers=none,
    columns=fullflexible,
}
\lstdefinestyle{customopencl}{%
    belowcaptionskip=1\baselineskip,
    breaklines=true,
    frame=none,
    xleftmargin=\parindent,
    language=Python,
    showstringspaces=false,
    basicstyle=\footnotesize\ttfamily,
    keywordstyle=\bfseries\color{red!40!black},
    commentstyle=\itshape\color{purple},
    identifierstyle=\color{black},
    stringstyle=\color{green!60!black},
    keywords=[2]{as,True,False,int,float},
    keywordstyle=[2]\bfseries\color{green!40!black},
    keywords=[3]{np,sp,plt},
    keywordstyle=[3]\bfseries\color{blue},
    keywords=[4]{0,1,2,3,4,5,6,7,8,9},
    keywordstyle=[4]\bfseries\color{magenta!80!black},
    columns=fullflexible,
}

\newcommand{\featuremem}[7]{\ifmmode f\text{-#1}^{\{#2,#3\}\{#4,#5\}}_{<#6>[#7]}\else $f\text{-#1}^{\{#2,#3\}\{#4,#5\}}_{<#6>[#7]}$\fi}

\newcommand{\featurememtag}[2]{\ifmmode f\text{-#1}_{(\text{#2})}\else $f\text{-#1}_{(\text{#2})}$\fi}

\newcommand{\featureop}[2]{\ifmmode f\text{-#1}_{<#2>}\else $f\text{-#1}_{<#2>}$\fi}

\newcommand{\featuresimp}[1]{\ifmmode f\text{-#1}\else $f\text{-#1}$\fi}

\newcommand{\kernelmem}[7]{\ifmmode k\text{-#1}^{\{#2,#3\}\{#4,#5\}}_{<#6>[#7]}\else $k\text{-#1}^{\{#2,#3\}\{#4,#5\}}_{<#6>[#7]}$\fi}

\newcommand{\kernelmemtag}[2]{\ifmmode k\text{-#1}_{(\text{#2})}\else $k\text{-#1}_{(\text{#2})}$\fi}

\newcommand\ldotscompact{\hbox to 1em{.\hss.\hss.}}

\newlength{\defaulttabcolsep}
\begin{document}

\runninghead{Stevens, Klöckner}

\title{A mechanism for balancing accuracy and scope in cross-machine
black-box GPU performance modeling}

\author{James D. Stevens\affilnum{1},
Andreas Klöckner\affilnum{1}}

\affiliation{\affilnum{1}University of Illinois at Urbana-Champaign; Deptartment of Computer Science}
\corrauth{James D. Stevens; 201 N Goodwin Ave; Urbana, IL 61801}
\email{jdsteve2@illinois.edu}



\begin{abstract}

The ability to model, analyze, and predict execution time of computations is an
important building block that supports numerous efforts, such as load balancing,
benchmarking, job scheduling, developer-guided performance optimization, and the
automation of performance tuning for high performance, parallel applications. In
today's increasingly heterogeneous computing environment, this task must be
accomplished efficiently across multiple architectures, including massively
parallel coprocessors like GPUs, which are increasingly prevalent in the world's
fastest supercomputers. To address this challenge, we present an approach for
constructing customizable, cross-machine performance models for GPU kernels,
including a mechanism to automatically and symbolically gather
performance-relevant kernel operation counts, a tool for formulating
mathematical models using these counts, and a customizable parameterized
collection of benchmark kernels used to calibrate models to GPUs in a black-box
fashion. With this approach, we empower the user to manage trade-offs between
model accuracy, evaluation speed, and generalizability. A user can define their
own model and customize the calibration process, making it as simple or complex
as desired, and as application-targeted or general as desired. As application
examples of our approach, we demonstrate both linear and nonlinear models;
these examples are designed to predict execution times for multiple variants of
a particular computation: two matrix-matrix multiplication variants, four
Discontinuous Galerkin (DG) differentiation operation variants, and two 2-D
five-point finite difference stencil variants. For each variant, we present
accuracy results on GPUs from multiple vendors and hardware generations. We
view this highly user-customizable approach as a response to a central question
arising in GPU performance modeling: how can we model GPU performance in a
cost-explanatory fashion while maintaining accuracy, evaluation speed,
portability, and ease of use, an attribute we believe precludes approaches
requiring manual collection of kernel or hardware statistics.


\end{abstract}

\keywords{Performance model, GPU, Microbenchmark, Code generation, Black box,
OpenCL}

\maketitle


\section{Introduction}%
\label{sec:intro}

Maximizing computational performance requires tailoring an application
implementation to the target architecture. As a result, obtaining and
maintaining good performance in a heterogeneous computing environment
necessitates the ability to efficiently decide between multiple mathematically
equivalent program variants. Being able to model, interpret, and predict the
execution time of computational kernels can provide insight into factors
contributing to computation cost, and doing so in an automated,
architecture-independent fashion is a key step toward the automation of
performance tuning for complicated, modern, vector-based, massively parallel
processor architectures including recent CPUs and GPUs.

GPUs, originally designed for rapid graphics rendering, have highly parallel
single instruction, multiple thread (SIMT) architectures that make them
particularly useful for data-parallel problems. Over the last
decade, general purpose GPU programming has risen in popularity. GPUs are
increasingly prevalent in the world's fastest supercomputers, and are being
utilized in an expanding body of applications including machine learning and
artificial intelligence. GPU programming has been facilitated by the release of
general purpose GPU programming systems, including Nvidia CUDA in 2007 and the
Open Computing Language (OpenCL) in 2009 \citep{nickolls2008cuda,munshi2011opencl}.

Tailoring a performance model to a particular computation on a single
hardware device may yield high accuracy. When broadening the scope of
architectures and computations targeted by a model, achieving high accuracy
becomes increasingly difficult. We present a mechanism for putting the user in
control of this trade-off between model accuracy and generalizability with an
approach for creating custom performance models that is realized on top of,
though technically not dependent on, a program transformation system.

We consider this contribution a building block to provide guidance in exploring
the vast search space of possible and, from the point of view of the result,
equivalent program variants, by either a developer or an auto-tuning compiler. We
view the models constructed within our framework as more economical alternatives
to evaluating execution time of computational kernels than, for example, using
actual on-device timing runs. Our system primarily targets execution on modern
GPU hardware, as exposed in, for example, the CUDA or OpenCL compute
abstractions. To facilitate portability and maximize ease of use, the system
makes few assumptions about the internal organization of the hardware, and
device-specific parameters are obtained from a black-box model-calibration
process that needs to run precisely once per model per hardware device used. We
demonstrate that this cross-machine, black-box, microbenchmarking approach to
analytical performance modeling can predict kernel execution time well enough
to determine which of multiple implementation variants will yield the shortest
execution times.

We review related performance modeling work in Section~\ref{sec:related_work},
and discuss two recent surveys of the current GPU performance modeling
landscape, \cite{madougou2016landscape} and \cite{lopez-novoa2015survey}. Like
the survey authors, we find that many existing GPU performance models predict
well for a particular application or architecture but are not easily portable
across machines. Most require knowledge of hardware or application
characteristics, often gathered manually, and significant effort to construct
and use. Compared to analytical approaches, learning and statistical techniques
tend to be more hardware-flexible, but the models produced are less accessible
to users from the standpoints of both design and interpretability; assumptions
and limitations about model predictive power, fidelity, and range of applicable
programs and hardware tend to be less clear. \cite{madougou2016landscape}
conclude that software utilizing these methods may be difficult to use for
users without good knowledge of statistical methods. None of the approaches we
surveyed provide users any significant control over the model expression or
(micro)benchmark design.

The following combination of factors distinguishes our work from previous GPU
performance modeling work, and comprises our primary contributions:

\begin{itemize}
\item Our approach allows broad customization of the mathematical model
    not available in previous work. A user can rely on predefined generic
    models or define their own model, making it as simple or as complex as
    desired, as application-targeted or as general-purpose as desired.
\item Similarly, our approach allows for complete customization of the set of
    measurement computations used to compute the model parameters during the
    model calibration process.
\item We automate the gathering of all performance-relevant kernel features used
    to model execution time. These features are gathered before compilation by
    examining our polyhedrally-based program representation. 
\item In the example models we show in Section~\ref{sec:results}, the only
    hardware statistic required is the \emph{sub-group size}, which is 32 on
    all current Nvidia and AMD hardware generations. This demonstrates that a
    black-box approach to performance modeling can capture execution cost
    behavior with very limited explicit representation of hardware
    characteristics. (Sub-groups and other details of the OpenCL machine model
    are discussed in Section~\ref{sec:opencl}.)
\item Models created using our approach are hardware vendor- and
    generation-independent, and we demonstrate performance on an AMD GPU and
    four generations of Nvidia GPUs.
\item Models created using our approach are amenable to human understanding:
    through the exposed parameters and their known meanings, it becomes
    possible to reason about which parts of kernel execution cost are
    attributed to which specific operations.
\item By making use of a polyhedrally-based program representation, we
    obtain precise counts of various units of work performed by the
    static-control part of a program. The counts obtained are parametric
    in the problem size, allowing us to amortize counting work over
    repeated applications of the model to the same kernels with varying
    problem size.
\item To help identify and measure cost contributions from individual
    memory accesses, we introduce a code transformation that can remove
    unrelated operations from a computation, thereby obtaining a kernel
    exercising the targeted memory access.
\item To help model temporal overlap, e.g., between computation and
    memory operations, we introduce a modeling paradigm that reflects the
    reduced apparent cost of the overlapped fraction of computation. We
    demonstrate the use of the resulting, complex model expressions within
    our (unmodified) black-box measurement and timing facilities to
    provide accurate performance modeling even in the challenging case of
    overlapped operations.
\end{itemize}

\subsection{Practical Realization of the Proposed Performance Modeling
Framework and the Loopy Program Transformation System}
\label{sec:interface}

For the representation of programs, as well as to enable our static counting of
operations, we rely on the \loopy\ program transformation system
\citep{kloeckner_loopy_2014,kloeckner_loopy_2015}.  While we make use of
certain capabilities available in \loopy, this is not a hard dependency, in
that any system offering a given interface could assume this role. In this
section, we first briefly introduce \loopy\ and then describe that abstract
interface.

\loopy\ is a programming system for array computations that targets CPUs, GPUs,
and other, potentially heterogeneous, compute architectures. This system keeps
the mathematical intent of a computation strictly separated from computational
and performance-related minutiae. To attain that goal, \loopy\ realizes
programs as objects in a host programming language (Python in this concrete
case) that can be manipulated from their initial, ``clean,'' mathematical
statement into highly device-specific, optimized versions via a broad array of
transformations. From these program objects, \loopy\ generates and executes
source code in a range of output languages, including OpenCL, the output
language we use for \loopy\ programs demonstrated in this work.

Our methodology to automatically gather the kernel statistics underlying
kernel features that are used by our modeling process leverages the
\loopy\ programming system in a number of ways:
\begin{itemize}
\item we express our kernels in its intermediate representation based
  on a generic OpenCL/CUDA-style machine model,
\item we use its program transformation vocabulary to obtain computationally
  different but mathematically equivalent variants of our measurement kernels,
\item we use it to generate OpenCL C-level source code for the various
  target machines on which we evaluate our model. While \loopy\ is able
  to target a much larger range of output languages (e.g., C, OpenMP [Open
  Multi-Processing] + SIMD [single instruction, multiple data], OpenCL, ISPC
  [Intel SPMD Program Compiler], CUDA), we limit ourselves to only OpenCL in
  keeping with our focus on GPUs, and finally,
\item we make use of \loopy's polyhedrally-based internal
  representation to support the automatic extraction of kernel statistics.
\end{itemize}

One of the primary design goals of \loopy, \perflex, and \uipick\ is to
avoid the need to write detailed OpenCL- or CUDA-level code manually,
thereby reducing development cost. We view this reduction of manual
kernel construction as a valuable feature. Nonetheless, our modeling approach
could be used with less reliance on \loopy, or none at all.

For example, the automated statistics-gathering piece of our work is
notionally independent of our modeling process in the sense that, while
it is convenient to have the ability to automatically extract the kernel
features being used in our models, it is not technically necessary and
could be achieved either by hand or in a technologically different
manner. One tool that could facilitate the collection of statistics to
compute feature values for a non-\loopy\ kernel is the Polyhedral
Extraction Tool, a library for extracting a polyhedral representation
from C source code \citep{verdoolaege2012polyhedral}.

Additionally, the process of predicting performance via model evaluation
can be easily separated from the model construction and calibration process.
Once a model has been calibrated and parameter values obtained using \perflex\
and \uipick, if a developer has a different technique to obtain statistics
and feature values for their application kernel, or if these values are
known by design, they may use the parameter values to predict performance
independently from \perflex\ and \loopy.

To accommodate legacy code in the existing system, we would also like to
highlight the capability of \loopy\ to ingest a dialect of Fortran 77, as
described in \cite{kloeckner_loopy_2015,kloeckner_loopy_2016}.




\subsection{OpenCL Machine Model}
\label{sec:opencl}

Program representation in \loopy\ relies on the OpenCL machine
model \citep{munshi2011opencl} for semantics, and typical GPU
hardware mappings thereof inspire the models we use to demonstrate our
approach. A very brief overview of the OpenCL machine model will introduce
terminology used throughout the following discussion.

The OpenCL model considers two levels of concurrency, each explicitly
exposed to the abstraction by the programmer in the form of a
multi-dimensional grid.  Each integer point in the grid is called a
\emph{work-item}. A rectangularly-indexed set of work-items forms a
\emph{work-group}, and a rectangularly-indexed set of work-groups forms a
\emph{grid} (or \emph{NDRange}), which is the unit in which work is
expressed to the abstraction. Additionally, a \emph{sub-group} is an
implementation-dependent grouping of work-items within a work-group which
we assume to approximate the effective vector width with which the
architecture issues instructions. Aside from barrier synchronization
across the work-items in a work-group, individual work items are assumed
to be independent. Item indices within a work-group are termed
\emph{local indices} or \emph{local IDs}, and indices of a work-group in
the grid are termed \emph{group indices} or \emph{group IDs}.  We will
use symbols like \texttt{lid(0)} to denote the local ID along axis 0 and
\texttt{gid(1)} to denote the group ID along axis 1. We assume, in
keeping with implementations available in the marketplace, that
work-groups are implemented by first mapping contiguous work-items along
the lowest-numbered axis to adjacent SIMD lanes, subsequently to
simultaneous multithreading (SMT), and finally to sequential execution.
Likewise we assume that work-groups are mapped to individual execution
cores, potentially mapping multiple work-groups onto one core depending
on capacity.

\section{Illustrative Example}
\label{sec:illustrative_example}

As an introduction to our modeling approach, consider the following simple,
illustrative example use case. Suppose a user wants to model and predict the
execution time of square matrix-matrix multiplication on a GPU. In this case,
we will only predict execution times for a single program variant employing an
algorithm which divides each matrix into $16 \times 16$ tiles and avoids some
repeated access to global memory by prefetching these tiles into local memory
shared between threads before performing the multiplication and addition.

\subsection{Kernel Creation and Transformation}
\label{sec:kernel_creation_and_transformation}

To construct an initial program representation of the matrix-matrix
multiplication workload using \loopy, we first specify the mathematical
intent:
\begin{tcolorbox}[listingbox]
\begin{lstlisting}[style=custompython]
knl = lp.make_kernel("{[i,j,k]: 0<=i,j,k<n}",
                       "c[i,j] = sum(k, a[i,k]*b[k,j])")
\end{lstlisting}
\end{tcolorbox}
\noindent
Without any code transformations, \loopy\ produces a sequential
algorithm looping over each index, as in the following OpenCL code:
\begin{tcolorbox}[listingbox]
\begin{lstlisting}[style=customopencl]
float acc;
for (int j = 0; j <= -1 + n; ++j)
  for (int i = 0; i <= -1 + n; ++i)
  {
    acc = 0.0f;
    for (int k = 0; k <= -1 + n; ++k)
      acc = acc + a[n*i + k] * b[n*k + j];
    c[n*i + j] = acc;
  }
\end{lstlisting}
\end{tcolorbox}
\noindent
To expose levels of concurrency in the form of loops that will become
parallelized, we perform loop-splitting transformations:
\begin{tcolorbox}[listingbox]
\begin{lstlisting}[style=custompython]
knl = lp.split_iname(knl, "i", 16)
knl = lp.split_iname(knl, "j", 16)
knl = lp.split_iname(knl, "k", 16)
knl = lp.assume(knl, "n >= 1 and n % 16 = 0")
\end{lstlisting}
\end{tcolorbox}
\noindent
Each \texttt{split\_iname} transformation divides one loop into two nested
loops with the inner loop iterating over the specified number of index values
(16). Without knowledge about the value of \texttt{n}, \loopy\ would need to
add several conditional statements to prevent out-of-bounds array access. We
avoid these conditionals by adding the \texttt{assume} transformation, yielding
the following OpenCL code:
\begin{tcolorbox}[listingbox]
\begin{lstlisting}[style=customopencl]
float acc;
for (int j_out = 0; j_out <= ...
  for (int j_in = 0; j_in <= 15; ++j_in)
    for (int i_out = 0; i_out <= ...
      for (int i_in = 0; i_in <= 15; ++i_in)
      {
        acc = 0.0f;
        for (int k_out = 0; k_out <= ...
          for (int k_in = 0; k_in <= 15; ++k_in)
            acc = acc +
                a[n*(16*i_out + i_in) + 16*k_out + k_in]
                *
                b[n*(16*k_out + k_in) + 16*j_out + j_in];
        c[n*(16*i_out + i_in) + 16*j_out + j_in] = acc;
      }
\end{lstlisting}
\end{tcolorbox}
\noindent
To parallelize loops, we tag loop indices with OpenCL thread indices:
\begin{tcolorbox}[listingbox]
\begin{lstlisting}[style=custompython]
knl = lp.tag_inames(knl,
    "i_out:g.1, i_in:l.1, j_out:g.0, j_in:l.0")
\end{lstlisting}
\end{tcolorbox}
\noindent
This \texttt{tag\_inames} transformation parallelizes loop(s) across thread
indices as specified, i.e., \texttt{"i\_in:l.1"} parallelizes the
\texttt{i\_in} loop across the \texttt{lid(1)} thread index. After tagging
loop indices, we obtain the following OpenCL code:
\begin{tcolorbox}[listingbox]
\begin{lstlisting}[style=customopencl]
float acc;
acc = 0.0f;
for (int k_out = 0; k_out <= ...
  for (int k_in = 0; k_in <= 15; ++k_in)
    acc = acc +
        a[n*(16*gid(1) + lid(1)) + 16*k_out + k_in] *
        b[n*(16*k_out + k_in) + 16*gid(0) + lid(0)];
c[n*(16*gid(1) + lid(1)) + 16*gid(0) + lid(0)] = acc;
\end{lstlisting}
\end{tcolorbox}
\noindent
Finally, to help amortize data motion cost, we perform prefetching
transformations:
\begin{tcolorbox}[listingbox]
\begin{lstlisting}[style=custompython]
knl = lp.add_prefetch(knl, "a", ["i_in","k_in"])
knl = lp.add_prefetch(knl, "b", ["k_in","j_in"])
\end{lstlisting}
\end{tcolorbox}
\noindent
The two prefetching transformations load data from the specified array into
local memory outside of the specified loop, yielding the following OpenCL
kernel:
\begin{tcolorbox}[listingbox]
\begin{lstlisting}[style=customopencl]
float acc;
acc = 0.0f;
__local float a_fetch[16*16];
__local float b_fetch[16*16];
for (int k_out = 0; k_out <= ((-16 + n) / 16); ++k_out)
{
  barrier(CLK_LOCAL_MEM_FENCE);
  a_fetch[16*lid(1) + lid(0)] =
      a[n*(16*gid(1) + lid(1)) + 16*k_out + lid(0)];
  b_fetch[16*lid(1) + lid(0)] =
      b[n*(16*k_out + lid(1)) + 16*gid(0) + lid(0)];
  barrier(CLK_LOCAL_MEM_FENCE);
  for (int k_in = 0; k_in <= 15; ++k_in)
    acc = acc + a_fetch[16*lid(1) + k_in] *
        b_fetch[16*k_in + lid(0)];
}
c[n*(16*gid(1) + lid(1)) + 16*gid(0) + lid(0)] = acc;
\end{lstlisting}
\end{tcolorbox}

\subsection{Model Creation, Calibration, and Evaluation}
\label{sec:model_creation_calibration_and_evaluation}

For simplicity in this example, we only predict execution times for a
single program variant, so a very simple mathematical model may suffice.
We model the execution time as
\begin{equation}
    t(n) \approx p_{\text{madd}}\cdot f_{\text{madd}}(n),
    \label{eq:example_model}
\end{equation}
where $n$ is matrix width, or, more generally a parameter which
statically determines the size of the loop domain that is assumed to be
known at runtime; $f_{\text{madd}}(n)$ is the number of multiply-add
(``madd'') sequences executed by the algorithm; $p_{\text{madd}}$ is a
hardware-dependent \emph{parameter} representing the effective cost
(seconds) per madd for this program variant; and $t(n)$ is execution
time. We distinguish madd sequences from isolated multiplication or
addition operations since many GPUs provide a specialized fused
multiply-add (FMA) operator. We refer to quantitative kernel characteristics
used in our models as kernel \emph{features}. \emph{Input features}, e.g.,
$f_{\text{madd}}(n)$, appear in the model function, and \emph{output features},
e.g., $t(n)$, are produced when evaluating a model function.

Our methods for automatically producing symbolic counts like
$f_{\text{madd}}(n)$ in the form of piecewise quasi-polynomials will be
discussed in Section~\ref{sec:stats}. Now we need a value for parameter
$p_{\text{madd}}$.

To determine this parameter value while treating the GPU as a black-box, we run
measurement computations designed to reveal the appropriate cost. In this
simple example we need the \emph{effective} madd cost only for \emph{this}
particular program variant, so we determine this cost by running the same
matrix multiplication variant on various sizes of matrices, computing
$f_{\text{madd}}(n)$ and $t(n)$ for each, and then calibrating our model by
fitting the model function to the feature data to compute $p_{\text{madd}}$.
More complex models that employ a microbenchmark approach to determine
parameter values will be discussed in Section~\ref{sec:results}.

The following code fragments demonstrate how this is accomplished using our
framework.

\medskip
1. Define the model shown in \eqref{eq:example_model} by specifying
wall time on the \nvidiamaxwell\ GPU as the output feature
(\texttt{f\_cl\_wall\_time\_nvidia\_geforce}), and providing the model
expression.  \texttt{p\_f32madd} represents parameter $p_{\text{madd}}$ and
\texttt{f\_op\_float32\_madd} specifies input feature $f_{\text{madd}}(n)$:
\begin{tcolorbox}[listingbox]
\begin{lstlisting}[style=custompython]
model = Model("f_cl_wall_time_nvidia_geforce",
                "p_f32madd * f_op_float32_madd")
\end{lstlisting}
\end{tcolorbox}

2. Generate measurement kernels for model calibration using our parameterized
collection of \emph{kernel generators}, \uipick, which we describe in
Section~\ref{sec:uipick}. To govern which generators are used and which kernel
variants are produced by each generator, we pass \emph{filtering tags} to the
kernel collection identifying the desired computation and describing, for
example, the data type of the arrays, the work-group dimensions, whether
to prefetch data into local memory, or whether to assume that
work-group dimensions divide evenly into the corresponding array dimensions. In
this example, we generate square matrix multiplication kernel variants:
\begin{tcolorbox}[listingbox]
\begin{lstlisting}[style=custompython]
filter_tags = [
    "matmul_sq", "dtype:float32", "prefetch:True",
    "lsize_0:16", "lsize_1:16", "groups_fit:True",
    "n:2048,2560,3072,3584"]
m_knls = KernelCollection(uipick.ALL_GENERATORS).generate_kernels(filter_tags)
\end{lstlisting}
\end{tcolorbox}

\noindent
\emph{Generator filter tags}, consisting of a single value, e.g.,
\texttt{matmul\_sq}, determine which generators run. This kernel collection
includes all \uipick\ generators, but the \texttt{matmul\_sq} tag filters out
generators not executing a square matmul. In this case, only generators with a
tag matching the user-supplied generator tag, \texttt{matmul\_sq}, will execute,
which leaves us with a single generator. \emph{Variant filter tags} consist of
\texttt{argument:value} pairs, e.g., \texttt{dtype:float32}, and determine which
kernel implementation variants are produced by the generators. Each generator
maintains a set of allowable values for each argument and generates one kernel
for each set of arguments in the Cartesian product of allowable argument value
sets. By passing a variant filter tag with argument values, the user can reduce
the set of allowable values for that argument. In this example, we provide a
set of four values for the problem size \texttt{n} and a single value for each
of the remaining arguments, so four measurement kernels will be generated. We
discuss kernel set generation and filtering further in
Section~\ref{sec:uipick}.

\medskip
3. Compute input and output feature values for all measurement kernels:
\begin{tcolorbox}[listingbox]
\begin{lstlisting}[style=custompython]
m_knl_feature_values = gather_feature_values(
    model.all_features(), m_knls)
\end{lstlisting}
\end{tcolorbox}
\noindent
Here we compute the two feature values found in our model, madd count and
execution time, for each of the four measurement kernels. This uses our
automatic kernel statistics gathering techniques described in
Section~\ref{sec:stats}.

4. Fit model to feature value data, producing parameter values:
\begin{tcolorbox}[listingbox]
\begin{lstlisting}[style=custompython]
model_param_values = fit_model(
    model, m_knl_feature_values)
\end{lstlisting}
\end{tcolorbox}
\noindent
We describe this calibration process in Section~\ref{sec:computing_params}.

5. Evaluate the model to predict execution time for a kernel:
\begin{tcolorbox}[listingbox]
\begin{lstlisting}[style=custompython]
exec_time = model.eval_with_kernel(
    model_param_values, test_knl, {"n": 1024})
\end{lstlisting}
\end{tcolorbox}

\begin{figure}
\begin{center}
    \includegraphics[width=0.48\textwidth]{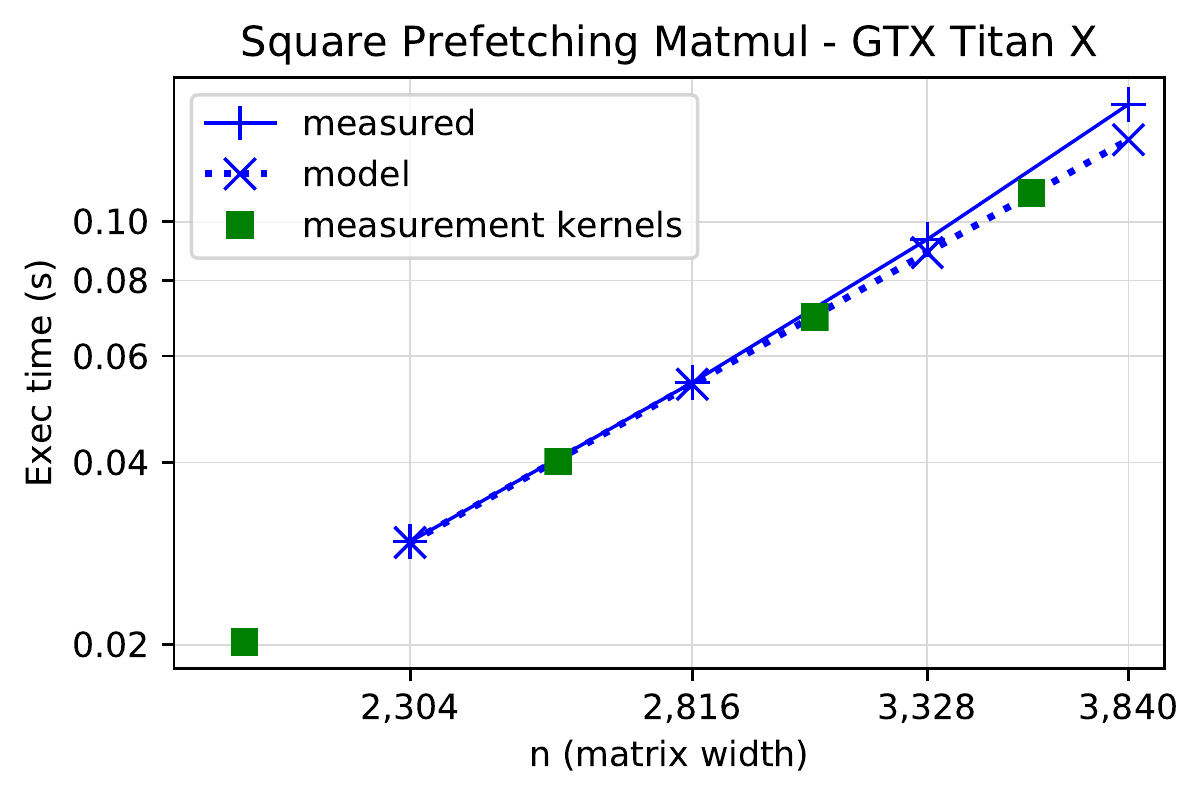}
    \caption{Measured vs. modeled execution times for square tiled matrix
    multiplication with prefetching on the \nvidiamaxwell\ GPU.
    \small{OpenCL/platform/driver info: OpenCL 1.2 CUDA 10.1.120 (430.50)}}
    \label{fig:results_illustrative_example}
\end{center}
\end{figure}

Figure~\ref{fig:results_illustrative_example} displays the measured execution
times and model predictions obtained in this example on an \nvidiamaxwell\
GPU.
In this case, we
sacrifice breadth of model applicability to achieve very accurate predictions
with a very simple model by using measurement kernels that are very similar to
our target computation.

We can achieve different modeling results, and further insight, by modifying
our measurement kernel set. For example, if we would like to explore the
component of execution time attributable to madd operations only, we can
replace the matrix multiplication kernels in our measurement set with kernels
designed to measure peak madd throughput by using the following filter tags: 
\begin{tcolorbox}[listingbox]
\begin{lstlisting}[style=custompython]
filter_tags = [
    "flops_madd_pattern", "dtype:float32",
    "lsize_0:16", "lsize_1:16", "groups_fit:True",
    "lid_stride_0:1", "lid_stride_1:2048",
    "nelements:524288,786432,1048576,1310720",
    "m:1024,1152,1280,1408"]
\end{lstlisting}
\end{tcolorbox}
\noindent
We discuss further details of this process in Section~\ref{sec:uipick}.

\begin{figure}
\begin{center}
    \includegraphics[width=0.48\textwidth]{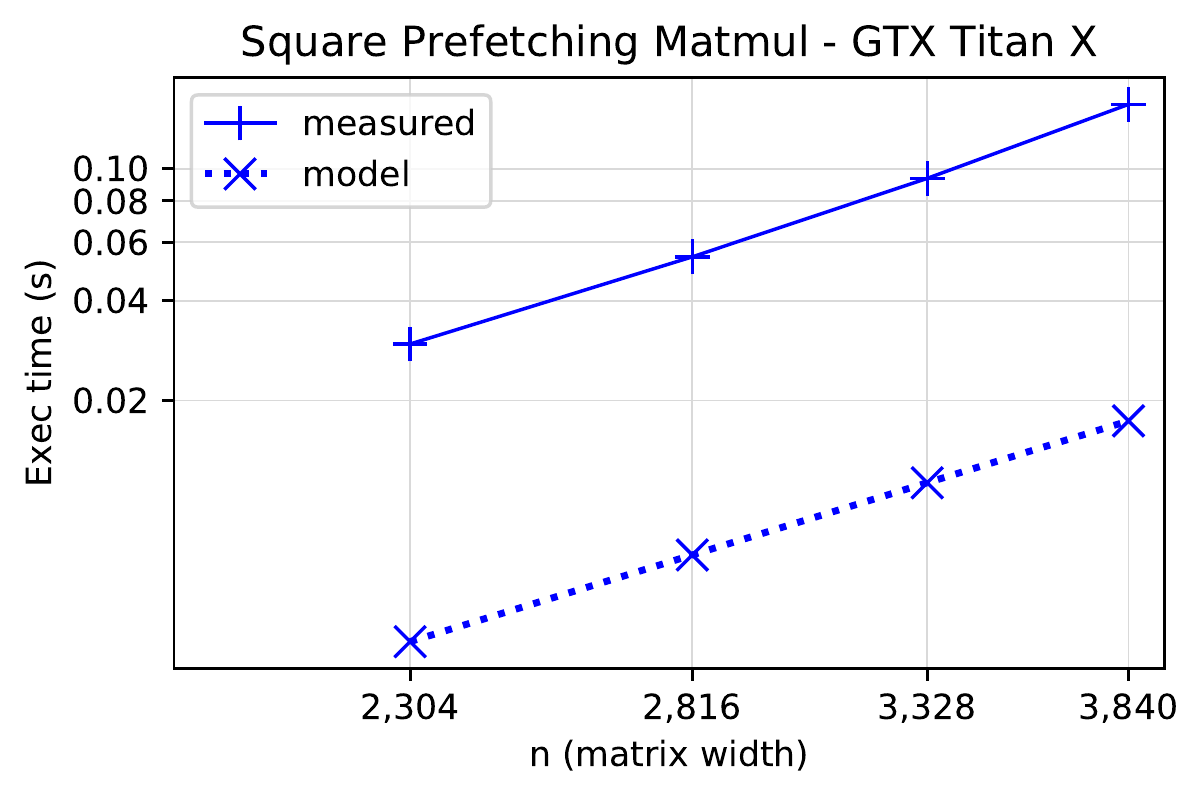}
    \caption{Modeling component of execution time attributable to madd
        operations for square tiled matrix multiplication with prefetching the
        \nvidiamaxwell\ GPU. \small{OpenCL/platform/driver info: OpenCL 1.2
        CUDA 10.1.120 (430.50)}}
    \label{fig:results_illustrative_example_madd}
\end{center}
\end{figure}

Calibrating the model with this measurement kernel set yields the result
shown in Figure~\ref{fig:results_illustrative_example_madd}. Since the value
for $p_{\text{madd}}$ in this model was computed using data from measurement
kernels designed to reveal peak FLOP/s (floating-point operations per second)
rates for madd operations, these results may provide insight into the
portion of the computation attributable to madd operations. Examples in
Section~\ref{sec:results} will demonstrate more complex models that use large
sets of microbenchmark measurement kernels to determine larger numbers of
parameter values, and predict execution time for multiple kernel variants.

\section{Overview of the Contribution}
\label{sec:overview}

There are three main components to our framework. \uipick, our
parameterized collection of kernels, contains kernel generators that
produce a customized set of measurement computations used to calibrate model
parameters.
A statistics gathering module, which we have added to \loopy, enables the
automated extraction of kernel statistics.
\perflex, our performance modeling tool, enables
custom model construction from user-defined parameters and kernel
features, as well as model calibration and evaluation.

\begin{figure*}
    \centering
    \includegraphics[width=\textwidth]{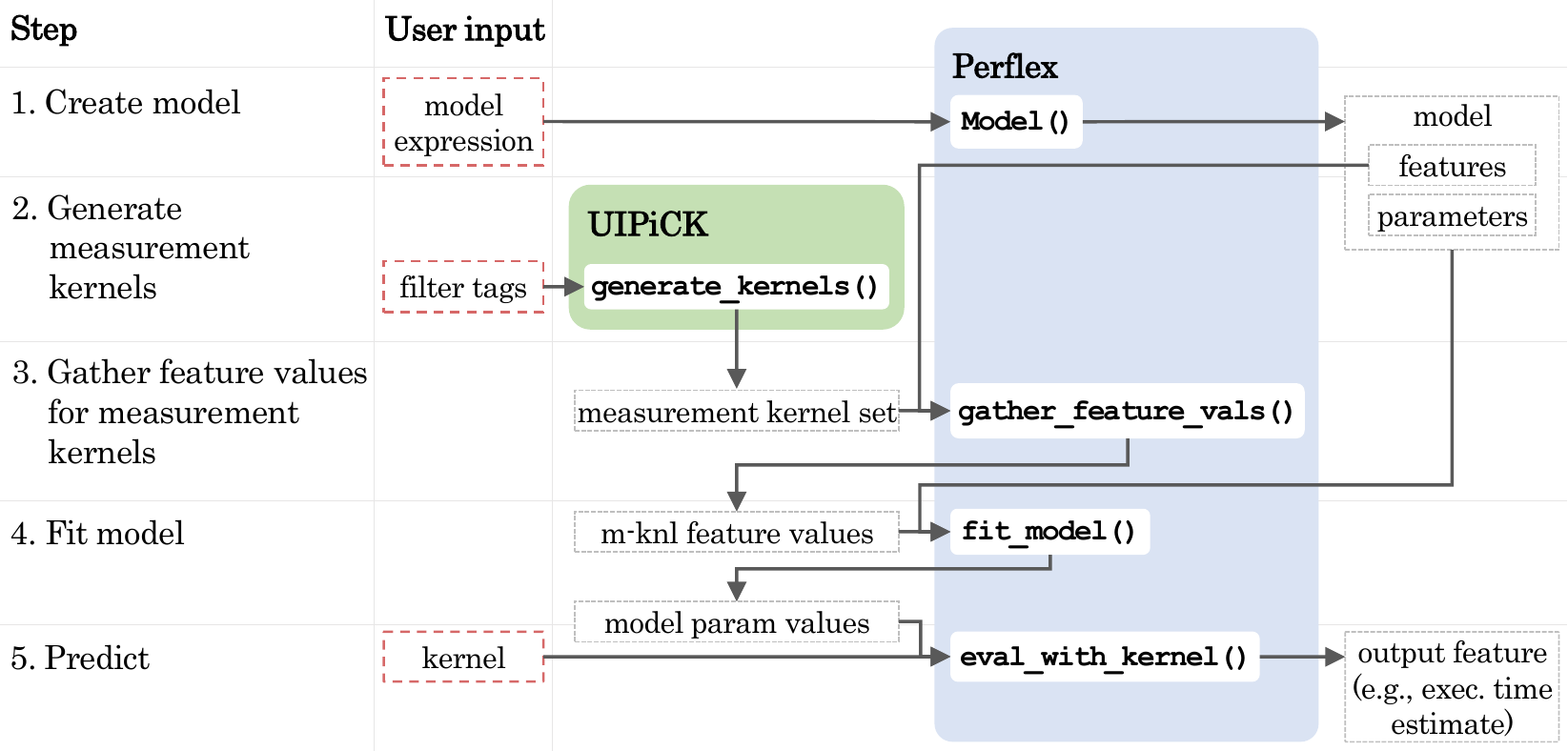}
    \caption{An overview of the performance modeling process.}
    \label{fig:process_overview}
\end{figure*}

Figure~\ref{fig:process_overview} shows the process for creating and calibrating a
model. First, the user creates a model as an arithmetic expression in terms of
some subset of the available \perflex\ features. Second, the user provides a
list of filtering tags to \uipick, which generates a measurement kernel set.
Then \perflex\ gathers feature values for each kernel in the set and calibrates the model by fitting the model function to the feature value data. This produces model parameter values, which
the user can then use, along with the model, to produce execution time
predictions for new kernels. \uipick\ generators use \loopy\ to create and
transform measurement kernels, and \perflex\ uses our statistics-gathering
routines when computing kernel feature values.

We organize this contribution as follows. In
Section~\ref{sec:assumptions}, we discuss limitations of our approach and
the assumptions underlying it. In Section~\ref{sec:stats}, we present our
procedure for automated extraction of performance-relevant kernel
statistics. In Section~\ref{sec:modeling}, we introduce a set of kernel
features derived from these statistics that form the basis of models
constructed using our framework. In
Section~\ref{sec:calibrating_model_parameters}, we describe how \uipick\
produces a customized set of measurement kernels based on user-provided
filtering tags, and \perflex\ enables custom model creation and calibration.
In Section~\ref{sec:results}, we evaluate the predictive power of several
performance models we create using \perflex\ and \uipick\ on five GPUs
from various hardware generations and vendors. Lastly, we compare our
approach to related work and provide some conclusions and directions for
future work.

\section{Assumptions and Limitations}%
\label{sec:assumptions}

The modeling system as presented has considerable generality and
extensibility, as the set of expressions usable for modeling is
unbounded in size, and the set of program features is user-extensible.
For evaluation of our system, we focus on
\emph{cost-explanatory} models of performance, i.e.\ models that
seek to explain execution time of a kernel as a sum of costs, each
of which is given by an empirically determined coefficient multiplied
by an operation count for a particular type of operation.
In keeping with this view, we make only very limited use of
nonlinearity. Importantly, this is a feature of our evaluation and our
preferred choice of model, not necessarily of the system itself.
It is true however that the preference for such models has biased the set
of kernel features that are built into the system.

We envision a number of use cases for our system. First, as shown in
Section~\ref{sec:illustrative_example}, it can be used to help a human
user put timing measurements into perspective and understand performance
characteristics of a computation on a given machine. Additionally, it can
serve as a pruning heuristic for an optimization procedure that considers
computationally (or even algorithmically) different
but mathematically equivalent versions of a given kernel.
The latter use case gives rise to our key criterion for evaluation, for
which we (here, manually) produce program variants and evaluate the
degree to which our model provides correct guidance for ranking the
execution time of the kernels presented to it. Succeeding at this task
would make the model an effective pruning strategy, as it would permit us
to efficiently rule out parts of an autotuning search space without
having to rely on execution of the actual program. This does not mean
that the tuning process \emph{must} be entirely execution-free. In fact,
as we demonstrate, important accuracy gains are available if we allow for
additional, on-line measurement runs that obtain calibration parameters
for features not thus far encountered.

As a secondary objective, we evaluate the accuracy with which our models
predict overall execution time, in terms of relative error. In a broad
overview of the literature, discussed in Section~\ref{sec:related_work},
no known performance model in the examined literature was able to
consistently attain better than single-digit percentages of relative
error on this task. Thus we evaluate our models to be roughly consistent
with this standard and view departures from it as a sign of potential
modeling issues.

Lastly, we evaluate the \emph{interpretability} of our models.  While, for our
stated purpose, we are most interested in the relative ranking of various
program variants, we choose execution time rather than ranking as our primary
model output. Ranking-based approaches (e.g.,~\cite{chen_learning_2018}) have
seen considerable success, however discerning contributions to their prediction
output poses a considerable challenge.  We only consider models that have a
simple, mostly linear form in terms of potential cost contributions. We define
interpretability of our models as the degree to which the calibrated model is
consistent with the cost-explanatory point of view. For example, models that
require negative weights are inconsistent with the notion of `cost', as carrying
out additional operations of any type should never result in a cost reduction.
Taken together, these factors enhance reliability and trustworthiness of our
methods.


A further aspect of ensuring reliability is a statement of the
conditions under which we can expect our models to satisfy the
criteria above. An important class of performance effects relate to
machine utilization. For example, the peak rate of floating point
operations feasible on a given machine varies with the number of cores
in use, or, analogously, the number of vector lanes used within
a subgroup/warp/wavefront/SIMD vector. The sizes of
available on-chip state space (register, scratchpad) may impact
utilization of scheduling slots, which in turn may impact the degree
to which latency may effectively be hidden. These are examples of
effects that our models do not (and cannot) account for. In return, the
elementary cost coefficients that our models obtain by the calibration process
are readily interpretable, e.g., by comparisons with the reciprocal of memory
bandwidth and peak FLOP/s rate.

In cases of less than full utilization, our models can still be used as
the computation in question remains in steady state, essentially
increasing the cost of a single operation to account for
less-than-optimal utilization. In cases of genuinely varying utilization,
the only viable path is to shrink the modeling granularity (e.g., to a
single core or SIMD lane) until the variation in utilization is no longer
relevant. We note that this is not at all an uncommon assumption. The
`execution-cache-memory' (ECM) family of
models~\citep{treibig_introducing_2010} is based around similar
considerations and calls this a `speed of light' assumption. As a
simplified proxy for this assumption, we enforce that nearly all of the
measurement kernels used to calibrate the models demonstrated here, as well as
the kernels whose execution time we model, use work-groups of size 256.

The set of features available by default in our modeling framework is the
source of a further set of limitations and assumptions that merit
discussion. The majority of these features are based on the detection of
a specific type of operation (e.g., a floating point operation, or a
specific type of memory access) and an accounting of the number of times
that the cost of the operation is incurred through repeated execution of
the site of the operation. To estimate the latter quantity, we assume
that the program under consideration exhibits static (i.e.,
non-data-dependent) control flow that is accurately represented by the
polyhedrally-given loop domain. Data-dependent loop control flow could in
principle be handled by allowing user-supplied `average' trip counts, but
we defer this to future work. Further, any conditionals present in the
code are accounted for by summing the cost contribution of both branches,
matching the cost behavior of GPUs under divergent control flow. All of
our kernel generators accept tags allowing a user to specify assumptions,
e.g., that the relationship between a particular index bound and relevant
work-group dimensions eliminates the need for conditionals that would
otherwise be necessary to avoid out-of-bounds array access. We use these
to minimize the number of conditionals in the measurement kernels and
test kernels.

The cost of memory access is particularly challenging to predict.
Effects that may contribute to this circumstance include the interaction
of caches and locality (statement-to-statement as well as vector), and
contention in the setting of banked memory. We have further observed that
array sizes above a certain threshold (e.g., 1~GiB on our AMD hardware)
can severely impact performance. We make no attempt to model the
implementation details of each target machine's memory subsystem.
Instead, we take a two-pronged approach
(Section~\ref{sec:data_motion_features}). For simple types of memory
access whose cost we expect to generalize across programs, we offer
descriptive classification by, e.g., inter-lane stride, utilization ratio,
and data width. Notably, our ability to determine these is predicated on the
(multi-dimensional) array subscript being quasi-affine, as they rely on
polyhedrally-based reasoning. For more complex access patterns, we permit
the memory access cost to be measured by executing the memory access in
isolation including its loop environment, retaining an additive accounting
of cost. Our mechanism enabling this approach will be discussed further in
Section~\ref{sec:work_remover}.

Lastly, while the present implementation of the proposed system is based
on OpenCL to attain vendor coverage, it would be straightforward to
realize an analogous system for the Nvidia CUDA compute abstraction
\citep{nickolls2008cuda}, particularly given that Loopy is capable of
generating CUDA code.



\section{Gathering Kernel Statistics}
\label{sec:stats}

The basic mathematical primitive underpinning our data gathering strategy is
the ability to count the number of integer points in a subset of the
$d$-dimensional integer tuples $\mathbb Z^d$ specified by affine
inequalities connected in disjunctive normal form (i.e., a disjunction of
conjunctions of affine inequalities). The output of this operation is a
piecewise quasi-polynomial in terms of problem size parameters that may
occur as part of the specification of the set of integers. E.g., the number
of integer points \texttt{(i,j)} in \texttt{\{ p<=i<n and p<=j<i+1 \}} is
\texttt{1/2*(n\textasciicircum2 + p\textasciicircum2 - 2*n*p + n - p)}. We
make use of \texttt{barvinok} library in conjunction with the \texttt{isl}
library \citep{verdoolaege_counting_2007,verdoolaege_isl_2010} to perform
this operation, with a fallback to a less accurate, simpler counting
technique that is used should \texttt{barvinok} not be available.
\texttt{barvinok} in turn is based on Barvinok's algorithm
\citep{barvinok1994polynomial}.

To obtain a count of a given type of operation, e.g., a certain kind of
memory access, we proceed as in Algorithm~\ref{alg:opcount}.

\begin{algorithm}
  \caption{Determine per-kernel count of per-statement operation.}
  \begin{algorithmic}
    \FOR{each statement $S$ in the kernel}
      \STATE Compute projection $\pi_S(D)$ of loop domain $D$ onto\\
        \hspace{\algorithmicindent} set of loop indices in which $S$ resides.
      \STATE Obtain symbolic count $|\pi_S(D)|$ of integer points in\\
        \hspace{\algorithmicindent}projection (piecewise quasi-polynomial represent-\\
        \hspace{\algorithmicindent}ing the number of times $S$ will be executed).
        \STATE Count operations ($n_{\text{ops},S}$) occurring in single\\
        \hspace{\algorithmicindent}instance of statement $S$ (e.g.\ by traversing left-\\
        \hspace{\algorithmicindent}and right-hand-side expressions).
    \ENDFOR
    \STATE Find the overall count for the desired operation as
    \begin{equation}
      n_{\text{ops}} = \sum_{\text{Statement $S$}} |\pi_S(D)| \cdot
      n_{\text{ops}, S}.
      \label{eq:opcount}
    \end{equation}
  \end{algorithmic}
  \label{alg:opcount}
\end{algorithm}

Some counting operations require ancillary processing. For instance,
determining the number of floating point operations or memory transactions of a
certain data type requires knowing the result data type, which is provided by a
type inference pass. We also identify multiply-add sequences in expression
trees since some processors support a fused multiply-add operation. When
counting global memory accesses, we track the group-ID and local-ID stride
components, which we obtain via an analysis of the array index components.
Counting (typically integer) arithmetic operations within array indices is
optional.

Operations counted using Algorithm~\ref{alg:opcount} carry a \emph{count
granularity} specifying whether they are counted once per
\emph{work-item}, \emph{sub-group}, or \emph{work-group}. On-chip
operations, i.e., arithmetic and local memory access, are counted at
sub-group granularity, and global memory operations are counted per
work-item, with the exception of global memory accesses with
\texttt{lid(0)} stride 0 (multiple threads access the same memory
location), which we refer to as \emph{uniform} accesses and count on a
per-sub-group basis. This counting scheme necessitates the only
user-provided hardware statistic required by any part of our approach,
the sub-group size.

Practically speaking, many different operation counts are extracted at once and
maintained in a mapping of operation kinds to operation counts. The map keys
contain characteristics of each operation for later computation of the kernel
features discussed in Section~\ref{sec:kernel_features}, and the map values are
piecewise quasi-polynomials. All arithmetic in \eqref{eq:opcount} is
then carried through to the values of the mapping and performed symbolically
on the piecewise quasi-polynomials therein. Once these quasi-polynomial counts
are determined for a particular kernel, they can be cheaply reevaluated for
changed problem sizes, represented as domain and kernel parameters.

In addition to total operation counts, data motion features in \perflex\ models,
discussed in Section~\ref{sec:data_motion_features}, may specify the ratio of
the number of element accesses to the number of elements accessed (we will call
the latter measure the "size of the access footprint"). This footprint is found
as shown in Algorithm~\ref{alg:footcount}. The construction of the index map
$I_j$ and the computation of the union crucially rely on the polyhedral
primitives in our approach.

\begin{algorithm}
  \caption{Determine accessed index footprint $F_v\subset \mathbb
  Z^{d'}$ for variable $v$.}
  \begin{algorithmic}
    \STATE Let $v$ be a $d'$-dimensional array.
    \FOR{each statement $S$ in the kernel}
      \STATE Compute the projection $\pi_S(D)$ of the loop domain\\
        \hspace{\algorithmicindent}$D$ onto the set of loop indices in which $S$ resides.
      \FOR{each access $j$ to $v$ in statement $S$}
        \STATE Determine the multi-dimensional index mapping\\
          \hspace{\algorithmicindent}$I_{S,j}:\mathbb Z^{d}\to \mathbb N_0^{d'}$ that takes a tuple of loop\\
          \hspace{\algorithmicindent}variables to the accessed indices. For example,\\
          \hspace{\algorithmicindent}the access \texttt{a[2*k+1, l+1]} would have an index\\
          \hspace{\algorithmicindent}mapping of $I_{S,j}(k,l)=(2k-1, l+1)$.
      \ENDFOR
    \ENDFOR
    \STATE
    Find the overall accessed footprint as
    \begin{equation*}
        F_v= \bigcup_{\substack{\text{Statement $S$,}\\
                                \text{access $j$}}}
                     I_{S,j}(\pi_S(D)).
    \end{equation*}
  \end{algorithmic}
  \label{alg:footcount}
\end{algorithm}


We count local memory accesses just as we do global memory accesses. We
can acquire other statistics, like work-group sizes and counts, by simply
querying a \loopy\ kernel object, like work-group sizes and counts.


Counting barrier synchronizations requires yet another approach, as these
are not apparent in \loopy\ code without a \emph{program linearization}. The
program linearization is found automatically by a search procedure and
determines the ordering of statements and the nesting of loops, which
enables a subsequent procedure that determines synchronization locations.
Once a program linearization is obtained and barriers are placed, the counting
process proceeds much as above, using the program linearization information to
obtain the relevant set of loop indices on which to project. Unlike with
data movement and floating point operations, the resulting count is the
number of synchronizations encountered by a single work-item.

Aside from enabling automatic computation of kernel features used in
\perflex\ models, this operation-counting procedure is an independently
useful tool for code analysis and algorithm development.

\section{Modeling Kernel Execution Time}
\label{sec:modeling}

We model execution time, or more generally any feature, as a function of
user-defined parameters and other kernel features, i.e.,
\begin{align*}
    T_{\text{wall}}(\mathbf n) &= \text{feat}^{\text{out}}(\mathbf n)\\
        & \approx
        g\left(\text{feat}_0^{\text{in}}(\mathbf n), \mathellipsis,
               \text{feat}_j^{\text{in}}(\mathbf n),
        p_0, \mathellipsis, p_k\right),
\end{align*}
where $\mathbf n$ is a vector of integer parameters used in the loop
domain that remain constant throughout the computation,
$\text{feat}_i^{\text{in}}(\mathbf n)$ accounts for the number of units of a
particular characteristic of a kernel (e.g., the number of single-precision
32-bit floating point multiplications), $p_i$ is a machine-dependent
calibration parameter related to hardware behavior, and the model expression
$g$ is a function provided by a user that is differentiable with respect to the
parameters. When creating a \perflex\ model, the user provides an output
feature $\text{feat}^{\text{out}}(\mathbf n)$ and a \emph{model expression}.

\subsection{Kernel Features}
\label{sec:kernel_features}

A \emph{kernel feature} is a function that accepts a kernel and a set of
domain parameters $\mathbf n$ and returns a (real) number. An \emph{input
feature} is a feature that appears in a model expression, like
$f_{\text{madd}}(n)$ in the example model described by
\eqref{eq:example_model}, and an \emph{output feature} is a
feature produced when evaluating a model, such as execution time. Any feature
may serve in either role.

\perflex\ uses the operation counting approach described in
Section~\ref{sec:stats} to compute features, which are parameterized by the
domain parameters $\mathbf n$. Once a symbolic representation like
$f_{\text{madd}}(n)$ has been determined from a kernel expressed in \loopy's
internal representation, it can be cheaply reevaluated for changed values of
the domain parameter vector $\mathbf n$. We cache these symbolic
representations for quick reuse, and \perflex\ distinguishes between situations
where a cached symbolic expression can be immediately re-evaluated using a
changed $\mathbf n$, and situations where additional processing using the new
problem size parameters is required. For example, a feature may have a
characteristic specified using inequality constraints involving a size
parameter, e.g., a memory access with \texttt{lstrides:\{0:<n\}}. The symbolic
expression for the number of accesses with this characteristic may change if
$n$ changes, so a previously computed expression cannot be directly reused.

When creating a model in \perflex, the user specifies an output feature
and a model equation containing input features. Each feature is denoted
by an identifier beginning with the prefix \texttt{f\_} as shown in
Section~\ref{sec:model_creation_calibration_and_evaluation}. The first
section of the identifier determines the feature class, and the remainder
determines specific characteristics of that feature. For example, we
might expand our matmul model to incorporate two types of memory access
costs as follows:
\begin{tcolorbox}[listingbox]
\begin{lstlisting}[style=custompython]
model = Model("f_cl_wall_time_nvidia_geforce",
    "p_f32madd * f_op_float32_madd + "
    "p_f32l * f_mem_access_local_float32 + "
    "p_f32g * f_mem_access_global_float32")
\end{lstlisting}
\end{tcolorbox}
\noindent
This model now contains one operation-counting feature and two memory
access-counting features. We provide a built-in set of features, discussed in
the following sections, which we have found sufficient to accurately model
execution time in a variety of computations. \perflex\ users can also create
their own custom features.

\subsubsection{Data Motion Features}
\label{sec:data_motion_features}

For most types of computational kernels, data motion is the dominant cost. We
account for this with a family of \emph{memory access features}, each member
of which has a set of characteristics affecting its cost. We refer to these
characteristics collectively as a \emph{memory access pattern}; they include

\begin{itemize}
    \item the \emph{memory type}, e.g., local or global,
    \item the \emph{direction}, e.g., load or store,
    \item the \emph{size} of the data type accessed, e.g., 32-bit or 64-bit,
    \item the \emph{local} and \emph{global strides} along each thread axis in
        the array index, i.e., strides \texttt{gs0, gs1, ..., ls0, ls1, ...} in
        flattened array index \texttt{array[gs0*gid(0) + gs1*gid(1) + ... +
        ls0*lid(0) + ls1*lid(1) + ...]} with units equal to the size of the
        data type (recall that we assume these indices are affine),
    \item and the \emph{ratio of the number of element accesses to the number
        of elements accessed} (access-to-footprint ratio, or \emph{AFR}). I.e.,
        a ratio of 1 means every element in the footprint is accessed one time
        and a ratio greater than 1 means that some elements are accessed more
        than once.
\end{itemize}

\textbf{Model Fidelity for Data Motion Features:}
Changes to any aspect of a memory access pattern may affect cost, particularly
for global memory access, and for this reason, we create a unique kernel
feature with characteristics matching each global memory access pattern found
in the kernels whose execution time we model in Section~\ref{sec:results}. As
an extreme example of the effect of access pattern modification on cost,
consider the difference between the patterns for matrices \texttt{a} and
\texttt{b} in the example in
Section~\ref{sec:kernel_creation_and_transformation}:
\begin{tcolorbox}[listingbox]
\begin{lstlisting}[style=customopencl]
for (int k_out = 0; k_out <= ((-16 + n) / 16); ++k_out)
  ...
  a_fetch[...] = a[n*(16*gid(1) + lid(1)) + 16*k_out + lid(0)];
  b_fetch[...] = b[n*(16*k_out + lid(1)) + 16*gid(0) + lid(0)];
\end{lstlisting}
\end{tcolorbox}

\noindent
The local strides and AFRs are the same for both arrays, and the loop
variable strides and global strides are different, as shown in
Table~\ref{table:ab_pattern}.

\begin{table}
    \centering
    \resizebox{\linewidth}{!}{%
    \begin{tabular}{lllll}
        \toprule
        & & \textbf{Local} & & \textbf{Loop} \\
        \textbf{Array} & \textbf{Ratio} & \textbf{strides} & \textbf{Global strides} & \textbf{stride} \\
        \midrule
        \texttt{a} & \texttt{n/16} & \texttt{\{0:1, 1:n\}} & \texttt{\{0:\ 0, 1:n*16\}} & \texttt{16} \\
        \texttt{b} & \texttt{n/16} & \texttt{\{0:1, 1:n\}} & \texttt{\{0:16, 1:0\}} & \texttt{16*n} \\
        \bottomrule
    \end{tabular}
    }
    \caption{Global load patterns in tiled matmul with prefetching.}
    \label{table:ab_pattern}
\end{table}

Despite these similarities, we observe that the cost can differ significantly
between these two patterns. To compare the costs of these two reads from
memory, we create two microbenchmark kernels, each of which reads a global
array using an access pattern matching the \texttt{a} or \texttt{b} fetch
pattern. We choose array sizes large enough that overhead and other costs
\emph{not} associated with the read are negligible. When we run them on the
\nvidiamaxwell\ GPU varying matrix width $n$ from 2048 to 3584, we observe an
average cost per load for the \texttt{b} pattern kernel that is consistently
4-5 times higher than that of the \texttt{a} pattern kernel. (Further platform
information is available in Table~\ref{table:gpus}.)

This observation provides evidence that aspects of GPU execution not
discoverable to a black-box model with justifiable effort can strongly
influence cost of global memory access. Observe that these two accesses only
differ in their group-id stride (i.e. work-group to work-group stride) and the
stride of the sequential loop variable $k$. Analytical models would have to
account for many undocumented machine details (e.g., work-group scheduling,
memory system architecture) to account for these differences.

Since it is difficult to rule out the possibility that any change in a global
memory access pattern will affect execution cost on some hardware, we create
a unique kernel feature with characteristics matching each different global
memory access pattern found in the kernels whose execution time we model in
Section~\ref{sec:results}. With this approach, a universal model for all
kernels on all hardware based on kernel-level features like ours would need a
prohibitively large number of global memory access features and corresponding
measurement kernels. This motivates our decision to allow proxies of
``in-situ'' memory accesses to be included as features, which in turn motivates
our `work removal' code transformation, discussed in
Section~\ref{sec:work_remover}. This transformation facilitates generation of
microbenchmarks exercising memory accesses which match the access patterns
found in specific computations by stripping away unrelated portions of the
computation in an automated fashion.


\textbf{Specifying Data Motion Features in the Model:}
To facilitate the creation of these individualized memory access features, we
provide the option to identify data motion features using a unique \emph{memory
access tag} to match by name a particular memory access found in a kernel. To
illustrate the use of this approach, we tag the global loads when creating our
\loopy\ matmul kernel with the identifiers \texttt{aLD} and \texttt{bLD} as
follows,
\begin{tcolorbox}[listingbox]
\begin{lstlisting}[style=custompython]
knl = lp.make_kernel(
    "{[i,j,k]: 0<=i,j,k<n}",
    "c[i,j] = sum(k, a$aLD[i,k]*b$bLD[k,j])")
\end{lstlisting}
\end{tcolorbox}
\noindent
and then define features counting these loads as follows:
\begin{tcolorbox}[listingbox]
\begin{lstlisting}[style=custompython]
model = Model("f_cl_wall_time_nvidia_geforce",
    "p_f32madd * f_op_float32_madd + "
    "p_f32l * f_mem_access_local_float32 + "
    "p_f32ga * f_mem_access_tag:aLD + "
    "p_f32gb * f_mem_access_tag:bLD + "
    "p_f32gc * f_mem_access_global_float32_store")
\end{lstlisting}
\end{tcolorbox}
\noindent
As we will discuss in Section~\ref{sec:work_remover}, in addition to
identifying memory access features in the model, these tags allow our work
removal code transformation to selectively remove computations from a kernel in
order to create microbenchmarks exercising specific memory access patterns.

As a less target-kernel-specific option, we also offer the possibility to
characterize memory accesses for the creation of memory access features by
specifying properties of the access pattern. In this case, each access matching
the property criteria is included when computing the feature. Using this
approach, we could incorporate these matrix multiplication memory accesses into
our model as follows:
\begin{tcolorbox}[listingbox]
\begin{lstlisting}[style=custompython]
model = Model("f_cl_wall_time_nvidia_geforce",
    "p_f32madd * f_op_float32_madd + "
    "p_f32l * f_mem_access_local_float32 + "
    "p_f32ga * f_mem_access_global_float32_load_
        lstrides:{0:1;1:>15}_gstrides:{0:0}_afr:>1 + "
    "p_f32gb * f_mem_access_global_float32_load_
        lstrides:{0:1;1:>15}_gstrides:{0:16}_afr:>1 + "
    "p_f32gc * f_mem_access_global_float32_store")
\end{lstlisting}
\end{tcolorbox}

All fields after the \texttt{f\_mem\_access} prefix are optional, and, in the
current implementation, they must be provided in the following order:
\begin{tcolorbox}[listingbox]
\begin{lstlisting}[style=custompython]
"f_mem_access_tag:<mem access tag>_
 <mem type>_<data type>_<direction>_
 lstrides:{<local stride constraints>}_
 gstrides:{<global stride constraints>}_
 afr:<AFR constraint>"
\end{lstlisting}
\end{tcolorbox}
\noindent


\subsubsection{Arithmetic Operation Features}
\label{sec:flop_features}

While execution time for many computations is dominated by data movement,
arithmetic operations also contribute, sometimes significantly, to overall
execution time. We account for these costs with a family of features that count
arithmetic operations. Each operation is characterized by the \emph{operation
type}, e.g., addition, multiplication, or exponentiation, and the \emph{data
type}, e.g., float32 or float64. A 32-bit floating point multiplication
operation feature, for example, could be specified in a model string as
\texttt{f\_op\_float32\_mul}. The models we demonstrate in this work do not
include integer arithmetic features; in the kernel variants modeled, integer
arithmetic is only used in array index computation, a cost that can be reduced
to negligible levels by, e.g., the compiler performing common subexpression
elimination.

\subsubsection{Synchronization Features}
\label{sec:sync_features}

Local barriers in GPU kernels halt execution of every thread within a
work-group until all threads have reached the barrier, and can contribute to
execution time. Additionally, launching a kernel incurs a constant overhead
cost. We account for these costs with a family of \emph{synchronization
features}. Synchronization types include local barriers and kernel launches.
Recall that the statistics gathering module counts the number of
synchronizations encountered by a single work-item, so depending on how a user
intends to model execution, they may need to multiply a synchronization feature
like local barriers by, e.g., the number of work-groups, a feature discussed in
the next section. A user might incorporate synchronization features into this
model as follows:
\begin{tcolorbox}[listingbox]
\begin{lstlisting}[style=custompython]
model = Model("f_cl_wall_time_nvidia_geforce",
    "p_f32madd * f_op_float32_madd + "
    ...
    "p_barrier * f_sync_barrier_local * f_thread_groups + "
    "p_launch * f_sync_kernel_launch")
\end{lstlisting}
\end{tcolorbox}

\subsubsection{Other Features}
\label{sec:other_features}

We provide a few other built-in kernel features in \perflex. By executing OpenCL
kernels containing no instructions and varying the number of work-groups
launched, we learned that average execution time increases with the work-group
count. This was true on all five GPUs we tested, listed in Table~\ref{table:gpus}.
We allow \perflex\ models to account for this cost by providing a \emph{thread
groups feature}, the total work-group count. We also provide an \emph{OpenCL
wall time feature}, which accepts a \emph{platform}, e.g., \texttt{nvidia}, and
\emph{device}, e.g., \texttt{geforce}, and when evaluated, executes 60
trials of the kernel on the specified device to obtain an average wall time.
This feature is typically chosen as the output in our model expressions. We
measure kernel execution time excluding any host-device transfer of data.

Based on the feature examples provided in the preceding sections, a complete
model might be expressed as follows:
\begin{tcolorbox}[listingbox]
\begin{lstlisting}[style=custompython]
model = Model("f_cl_wall_time_nvidia_geforce",
    "p_f32madd * f_op_float32_madd + "
    "p_f32l * f_mem_access_local_float32 + "
    "p_f32ga * f_mem_access_global_float32_load_lstrides:{0:1;1:>15}_gstrides:{0:0}_afr:>1 + "
    "p_f32gb * f_mem_access_global_float32_load_lstrides:{0:1;1:>15}_gstrides:{0:16}_afr:>1 + "
    "p_f32gc * f_mem_access_global_float32_store + "
    "p_barrier * f_sync_barrier_local * f_thread_groups + "
    "p_group * f_thread_groups + "
    "p_launch * f_sync_kernel_launch")
\end{lstlisting}
\end{tcolorbox}

\subsection{Model Parameters}
\label{sec:model_params}

Feature values, as discussed in the previous sections, are dependent on the
kernel and the (often size-related) domain parameters, whereas parameter values
are hardware-dependent. For example, the parameter $p_{\text{madd}}$ in
\eqref{eq:example_model}, passed as \texttt{p\_f32madd} to the example
\perflex\ model in Section~\ref{sec:model_creation_calibration_and_evaluation},
is the coefficient of the madd count in the model, representing the effective
cost per madd.  Identifiers of parameters in model expressions begin with
prefix \texttt{p\_} followed by a unique user-defined string of characters used
to distinguish the parameter from others. We determine parameter values using
the calibration process described in
Section~\ref{sec:calibrating_model_parameters}.

\section{Calibrating Model Parameters}
\label{sec:calibrating_model_parameters}

To avoid the need for machine-specific architectural and performance
knowledge, and to promote model portability and customizablility, we treat
the GPU as a black box and determine values for model parameters by
executing a set of \emph{measurement kernels}. We provide a collection of
them in a software package called \uipick, which enables this customizable
microbenchmarking functionality.

\subsection{Parameterized Collection of Kernels}
\label{sec:uipick}

\uipick\ includes a collection of over 20 \emph{kernel creation functions},
each capable of producing multiple implementation variants of a particular
\loopy\ kernel. These include computations designed to exercise a particular
feature, e.g., single-precision floating point multiplication or a particular
memory access pattern, as well as more complex application-oriented
computations, e.g., multiplying two matrices or performing matrix
transposition. Arguments passed to a creation function determine which variant
of a particular computation is produced. Arguments may include, for example,
thread group dimensions, array dimensions, data type, or whether to perform a
particular \loopy\ transformation like prefetching or loop unrolling.

While a user can use kernel creation functions directly to produce \loopy\
kernels, \uipick\ also provides a tag-driven filtering interface to facilitate
production of large sets of measurement kernels matching specified
characteristics. To do this, we provide a collection of \emph{kernel
generators}, each of which corresponds to a kernel creation function. Each
generator maintains a collection of filtering tags of two varieties.

\emph{Generator filter tags} determine which generators are used and consist of
a single value that identifies a characteristic of the computation, e.g.,
\texttt{matmul\_sq} or \texttt{flops\_mul\_pattern}. The generators matching the
user-provided generator filter tags, according to a user-provided
\emph{generator match condition}, will execute. The match condition defines the
condition under which \uipick\ will consider a particular generator a match for
the set of generator filter tags. We provide four possible match conditions:
(1) a generator's filter tag set must be identical to the user-provided tags,
(2) a generator's filter tag set must be a subset of the user-provided tags,
(3) a generator's filter tag set must be a superset of the user-provided tags
(default), or (4) the intersection of a generator's filter tag set and the
user-provided tags must be non-empty. An example below briefly illustrates how
different generator match conditions may yield different sets of matching
generators. 

\emph{Variant filter tags} consist of \texttt{argument:value} pairs, e.g.,
\texttt{dtype:float32}, and determine specific characteristics of the kernel
variants to be generated. For each argument, a generator maintains criteria
defining a set of allowable values. For example, the \texttt{prefetch} argument
might allow the set $\{\texttt{True}, \texttt{False}\}$, and the \texttt{dtype}
argument might allow the set $\{\texttt{float32}, \texttt{float64}\}$. When
executed, a generator generates one kernel for each set of arguments in the
Cartesian product of allowable argument value sets. By passing a variant filter
tag with argument values, a user reduces the set of allowable values for that
argument to a subset.

For example, recall the filter tags used in the example in
Section~\ref{sec:model_creation_calibration_and_evaluation}:
\begin{tcolorbox}[listingbox]
\begin{lstlisting}[style=custompython]
filter_tags = [
    "matmul_sq", "dtype:float32", "prefetch:True",
    "lsize_0:16", "lsize_1:16", "groups_fit:True",
    "n:2048,2560,3072,3584"]
m_knls = KernelCollection(uipick.ALL_GENERATORS).generate_kernels(filter_tags)
\end{lstlisting}
\end{tcolorbox}
\noindent
In this case, we do not provide a generator match condition, so the default,
condition (3), will be used; in order to execute, a generator's filter tag set
must be a superset of the user-provided generator tags. We provide one generator filter
tag, \texttt{matmul\_sq}, and only one generator in \uipick\ contains this tag,
so only this generator meets the match condition. If we had provided a second
generator filter tag, e.g., \texttt{finite\_diff}, then no generators would
meet the match condition since no generator's tag set contains both
\texttt{matmul\_sq} and \texttt{finite\_diff}. If we wanted both of these
generators to execute using such a tag set, we could instead use match condition (4), the
intersection of a generator's filter tag set and the user-provided tags must be
non-empty. We would specify this condition by passing
\texttt{generator\_match\_cond=MatchCondition.INTERSECT} as an argument to
\texttt{KernelCollection}.

Only one generator matches our single generator tag, so only one generator will
execute. We also provide a single value for each variant filter tag associated
with that generator, except array size \texttt{n}, for which we provide a set
of four values. The Cartesian product of these sets contains four sets of
argument values. These four sets each contain a different value for \texttt{n}
and are otherwise identical, so all four kernels produced will perform the same
square matmul with a different value of \texttt{n}. These variants will all
operate on 32-bit floating point data, perform a prefetching operation that
fetches $16 \times 16$ tiles from global memory, and assume that work-group
dimensions fit evenly into array dimensions, which avoids the need for
conditionals. If we were to omit, for example, the tag \texttt{prefetch:True},
we would instead obtain 8 kernels; for each problem size \uipick\ would
generate one kernel that performs prefetching and one that does not. We could
include more generators by adding additional generator filter tags and passing
the desired generator match condition to our kernel collection.

\subsubsection{Measurement Workload Synthesis by Work Removal Transformation}
\label{sec:work_remover}

To enable kernel-specific data motion features (described in
Section~\ref{sec:data_motion_features}), including the creation of
microbenchmarks exercising these features, and potentially other fine-grained
study of contributions of various operation types to computational cost, we
introduce a code transformation that can extract a set of desired operations
from a given computation, while maintaining overall loop structure and
sufficient data flow to avoid elimination of further parts of the computation
by optimizing compilers. We call this facility the `work remover'. 

This code transformation strips away arithmetic operations and local memory
access, i.e., on-chip work, from a kernel. It leaves behind all global
memory accesses or a subset thereof, as specified by the user, helping a
developer target and analyze specific portions of an application. For example,
it can help reveal the extent to which on-chip work and global memory access
affect execution time and inform the decision of whether to use a model that
allows for overlap of on-chip and global memory operations, as we will discuss
in Section~\ref{sec:results_models}.

Additionally, this tool can aid in producing a microbenchmark matching a
particular memory access pattern. As we will discuss in
Section~\ref{sec:mkernel_design}, we provide a generator that automatically
constructs measurement kernels exercising access patterns meeting certain
criteria. For more complex patterns, we use generators employing a subtractive,
rather than additive, approach to microbenchmark creation. Using
Algorithm~\ref{alg:remove_work}, these generators remove statements from a
target kernel, leaving behind a measurement kernel exercising the desired
memory access pattern.

For example, consider our tiled matrix multiplication \loopy\ kernel that
produced the following OpenCL code:
\begin{tcolorbox}[listingbox]
\begin{lstlisting}[style=customopencl]
float acc;
acc = 0.0f;
__local float a_fetch[16*16];
__local float b_fetch[16*16];
for (int k_out = 0; k_out <= ((-16 + n) / 16); ++k_out)
{
  barrier(CLK_LOCAL_MEM_FENCE);
  a_fetch[16*lid(1) + lid(0)] =
      a[n*(16*gid(1) + lid(1)) + 16*k_out + lid(0)];
  b_fetch[16*lid(1) + lid(0)] =
      b[n*(16*k_out + lid(1)) + 16*gid(0) + lid(0)];
  barrier(CLK_LOCAL_MEM_FENCE);
  for (int k_in = 0; k_in <= 15; ++k_in)
    acc = acc + a_fetch[16*lid(1) + k_in] *
        b_fetch[16*k_in + lid(0)];
}
c[n*(16*gid(1) + lid(1)) + 16*gid(0) + lid(0)] = acc;
\end{lstlisting}
\end{tcolorbox}
\noindent
To isolate the global load from \texttt{b}, we can use the work remover as
follows:
\begin{tcolorbox}[listingbox]
\begin{lstlisting}[style=custompython]
knl = remove_work(knl, remove_vars=["a", "c"])
\end{lstlisting}
\end{tcolorbox}
\noindent
Following Algorithm~\ref{alg:remove_work}, this function removes the local
memory transactions, as well as global memory accesses to variables
\texttt{a} and \texttt{c}, producing the following OpenCL kernel:
\begin{tcolorbox}[listingbox]
\begin{lstlisting}[style=customopencl]
float read_tgt;
read_tgt = 0.0f;
for (int k_out = 0; k_out <= ((-16 + n) / 16); ++k_out)
  read_tgt = read_tgt +
      b[n*(16*k_out + lid(1)) + 16*gid(0) + lid(0)];
read_tgt_dest[16*n*gid(1) + n*lid(1) + 16*gid(0) + lid(0)] = read_tgt;
\end{lstlisting}
\end{tcolorbox}
\noindent
Observe that the access pattern to \texttt{b} is unchanged. We include the
seemingly unnecessary global store to \texttt{read\_tgt\_dest} to ensure
that statements with unused results are not dropped by an optimizing
compiler. To maximize model accuracy, this store will also need to be
represented by a feature in the model; it uses a straightforwardly modeled,
stride-1 access pattern.

The code resulting from this transformation includes a tight dependency chain
on the increment of \texttt{read\_tgt}. These dependencies have the potential
to impact the measurements, mitigated, however, by low register usage and
therefore likely good latency hiding. We leave further investigation of this
matter to future work.

\begin{algorithm}
  \caption{Remove arithmetic and local memory operations.}
  \begin{algorithmic}
    \REQUIRE Set $G_\text{rm}$: global mem. accesses to remove.
    \STATE Insert statement initializing local var. $\text{tgt}_\text{read}=0$.
    \FOR{each statement $S$ in kernel}
      \STATE Let $G_\text{ld}(S)$ be the set of global memory loads in $S$.
      \STATE Let $G_\text{ld}^\text{new}(S) = G_\text{ld}(S) - (G_\text{ld}(S) \cap G_\text{rm})$.
      \FOR{each memory access $g_\text{ld}^\text{new}$ in $G_\text{ld}^\text{new}(S)$}
        \STATE Insert statement $\text{tgt}_\text{read} = \text{tgt}_\text{read} + g_\text{ld}^\text{new}$.\\ 
        \small{(keep dependencies of $S$; add dep. on $\text{tgt}_\text{read}$ init.)}
      \ENDFOR
      \IF{$S$ contains a global store $g_\text{st} \notin G_\text{rm}$}
        \STATE Insert statement $g_\text{st} = \text{tgt}_\text{read}$.\\
        \small{(keep dependencies of $S$; add dep. on $\text{tgt}_\text{read}$ init.)}
      \ENDIF
      \STATE Remove statement $S$.
    \ENDFOR
    \IF{$\text{tgt}_\text{read}$ is never written to global memory}
      \STATE Create new global variable $\text{tgt}_\text{read}^\text{dest}$ with one entry\\
        \hspace{\algorithmicindent}per work-item.
      \STATE Insert statement $\text{tgt}_\text{read}^\text{dest} =
\text{tgt}_\text{read}$ after all state-\\
        \hspace{\algorithmicindent}ments that modify
$\text{tgt}_\text{read}$, writing $\text{tgt}_\text{read}$ to\\
        \hspace{\algorithmicindent}the local work-item's entry in
$\text{tgt}_\text{read}^\text{dest}$.
    \ENDIF
  \STATE Infer type for $\text{tgt}_\text{read}$ and $\text{tgt}_\text{read}^\text{dest}$ based on $G_\text{ld}^\text{new}(S)$
  \end{algorithmic}
  \label{alg:remove_work}
\end{algorithm}

\subsubsection{Measurement Kernel Design}
\label{sec:mkernel_design}

To allow developers to create custom sets of microbenchmarks on which to
calibrate their models, \uipick\ allows users to choose from generators that we
provide, and developers may also create custom measurement kernel generators to
fit a specific purpose. In support of this work, we have created a set of
microbenchmark measurement kernels, each designed to measure the cost of a
single operation (as represented by a feature, see
Section~\ref{sec:kernel_features}). These kernels make up the measurement
kernel sets used to calibrate the models demonstrated in
Section~\ref{sec:results}, where we describe a rigorous evaluation procedure to
characterize the achievable modeling fidelity of the proposed system. One
important technique in designing these kernel sets is to, e.g., vary the
quantity of a single feature between kernels while keeping other feature counts
constant. Descriptions of a few fundamental examples of the microbenchmarking
kernels follow.

\textbf{Global memory access:}
We employ two varieties of measurement kernels designed to exercise global
memory access. First, for access patterns with AFR equal to one that are 
simple enough to be fully specified by the local strides, global strides,
and data size, i.e., patterns that do not produce a write race and are not
nested inside sequential loops, we provide a generator that automatically
constructs measurement kernels exercising the desired pattern, which is
specified via the user-provided variant filter tags. In these kernels, each
work-item performs a global load from each of a variable number of input
arrays using the specified access pattern. Each work-item then stores the
sum of the input array values it fetched in a single result array. For
example, with two load arrays, this kernel would perform the operation
\texttt{result[pattern] = in0[pattern] + in1[pattern]}. Variant filter tags
provided for these kernels specify the data type, global memory array size,
work-group dimensions, number of input arrays, and thread index strides.

Second, to generate measurement kernels exercising more complex access
patterns, e.g., memory accesses inside nested loops or with AFRs not equal
to one, we provide generators that use the work removal approach described
in Section~\ref{sec:work_remover} to create microbenchmark kernels exactly
matching these more complex patterns. These generators first construct the
original application kernel that contains the desired memory access pattern,
and then, using the process described in Section~\ref{sec:work_remover},
strip away operations until the desired memory access remains. These
generators tag memory accesses in these kernels using \loopy\ \emph{memory
access tags} and we use these memory access tags to identify the desired
access pattern in the corresponding \perflex\ model feature, as shown in
Section~\ref{sec:data_motion_features}. Variant filter tags
provided to these kernel generators include all filter tags used when
specifying the original application kernel variant, as well as filter tags
specifying whether to remove on-chip work and filter tags specifying which,
if any, global memory accesses to remove. 

\textbf{Arithmetic operations:}
In measurement kernels designed to exercise an arithmetic operation, we
first have each work-item initialize 32 private variables of the specified
data type.  We then have it perform a loop in which each iteration updates
each variable using the target arithmetic operation on values from other
variables. We unroll the loop by a factor of 64 and arrange the variable
assignment order to achieve high throughput using the approach found in the
Scalable HeterOgeneous Computing (SHOC) OpenCL \texttt{MaxFlops.cpp}
benchmark \citep{danalis2010scalable}. In this approach, the 32 variable
updates are ordered so that no assignment depends on the most recent four
statements. In our experience, using 32 variables permits peak performance
to be attained by avoiding tight dependency chains without losing
performance due to underutilization of scheduler slots due to register
space consumption.

After the loop completes, we sum the 32 variable values and store the result in
a global array according to a user-specified memory access pattern. We include
these global stores to ensure that statements with unused results are not
dropped by an optimizing compiler. These generators accept global access
patterns of the simple variety described above for which we can construct
kernels in an automated fashion. Variant filter tags provided for these kernels
include the data type, global memory array size, work-group dimensions, thread
index strides for the global memory access pattern, and number of loop
iterations.


\textbf{Local memory access:}
In measurement kernels designed to exercise access to OpenCL \texttt{local}
memory, i.e., access to the per-core shared scratchpad, each work-item
initializes one element of a local array to the data type specified. We
then have it perform a loop, at each iteration moving a different element
from one location in the array to another. We avoid write-races and
simultaneous reads from a single memory location, and use an
\texttt{lid(0)} stride of 1, avoiding bank conflicts.  After the loop
completes, each work-item writes one value from the shared array to global
memory. While our framework allows local memory access features to be
characterized by thread index strides, we do not use these strides to
differentiate local memory accesses in the demonstrations presented here.
Instead, we use a single feature for all 32-bit local memory accesses
occurring in measurement kernels and modeled kernels.  Variant filter tags
provided for this kernel include the data type, global memory array size,
iteration count, and work-group dimensions, which determine the local
strides for local memory access as well as the size of the local array.

\textbf{Other features:}
When creating the measurement kernel sets used to calibrate the models that
will be demonstrated in Section~\ref{sec:results}, we also generate variants of
a measurement kernel that executes a variable number of local barriers, a
measurement kernel that reveals operation overlapping behavior using the
strategy that will be described in Section~\ref{sec:modeling_overlap}, and a
measurement kernel that launches a specified number of work-groups performing
no operations. We set problem sizes to attain execution times between 1 and
1000 milliseconds, with the exception of the empty kernel generator, which
produces some kernels launching as few as 16 work-groups in order to reveal the
kernel launch overhead. Using a sufficiently high-fidelity model, we expect
that users will be able to differentiate between latency-based costs of a
single kernel launch and throughput-related costs that would be incurred in
pipelined launches.

\subsection{Computing Model Parameters}
\label{sec:computing_params}

After creating a model and generating a measurement kernel set, we collect
feature values for each measurement kernel and then fit the model function to
the data by minimizing the Euclidean norm of the residual in the nonlinear
least squares problem
\begin{equation*}
    \min_\mathbf{p}||\mathbf{g}(\mathbf{p}) - \mathbf{t}||_2,
\end{equation*}
where the residual is defined as 
\begin{align*}
    \mathbf{r}(\mathbf{p}) &= \mathbf{t} - \mathbf{g}(\mathbf{p})
\end{align*}
with
\begin{equation*}
\begin{aligned}[c]
    r_k(\mathbf{p}) &= t_k - g_k(\mathbf{p})\ \ (k \in \{0, \ldots, l\}),\ \ \\
    \mathbf{p}&=(p_0, \ldots, p_m)^T,\ \ \ \ \ \ \ \text{and }
\end{aligned}
\begin{aligned}[c]
    \mathbf{g}&=(g_0, \ldots, g_l)^T,\\
    \mathbf{t}&=(t_0, \ldots, t_l)^T.
\end{aligned}
\end{equation*}
Here, $l$ is the number of measurement kernels, $m$ is the number of model
parameters, $p_i$ is the $i$\textsuperscript{th} model parameter, $g_k$ is
the model function containing feature values for the
$k$\textsuperscript{th} measurement kernel, and $t_k$ is the output feature
value for the $k$\textsuperscript{th} measurement kernel. Thus, the
resulting nonlinear system contains one row for each measurement kernel.
Solving this problem involves evaluating the Jacobian:
\begin{equation*}
    \mathbf{J}(\mathbf{p}) : j_{ki} = \frac{\partial g_k}{\partial p_i}(\mathbf{p})
\end{equation*}
\noindent
After using symbolic differentiation to obtain the Jacobian, we provide the
Jacobian and residual to \textsc{Scipy}'s \texttt{optimize.leastsq} function,
which solves the nonlinear system using the Levenberg-Marquardt method.

If the user is concerned about prediction error relative to execution time,
rather than absolute prediction error, they may call
\texttt{scale\_features\_by\_output()} on the feature data before calibrating
the model, which divides each input feature value by the corresponding output
feature value and sets output feature values to 1. We perform this scaling
in all examples discussed in this work.

After calibrating the model, \perflex\ logs the least squares residual (as
defined above) evaluated at the solution. This value can be examined by a user
and may aid in assessing their model expression; if calibrating a particular
model to a particular workload results in a high residual, this may indicate
that the model does not fit the workload behavior well. We discuss other
strategies to determine the appropriateness of a model in
Section~\ref{sec:results_models}.

\subsection{Predicting Cost}
\label{sec:predicting_cost}
After obtaining model parameter values using the calibration process just
described, we can compute a predicted output feature (e.g., execution time)
for a \loopy\ kernel. This requires a dictionary of kernel arguments, i.e.,
problem size variable values, to compute the kernel feature values, which
are used along with the parameter values to evaluate the model as
demonstrated in the example in
Section~\ref{sec:model_creation_calibration_and_evaluation}:
\begin{tcolorbox}[listingbox]
\begin{lstlisting}[style=custompython]
model_param_values = fit_model(
    model, m_knl_feature_values)
exec_time = model.eval_with_kernel(
    model_param_values, test_knl, {"n": 1024})
\end{lstlisting}
\end{tcolorbox}
\noindent
Model evaluation cost is primarily driven by the evaluation of piecewise
quasi-polynomials as well as the model expression. 

\subsection{Modeling Operation Overlap}
\label{sec:modeling_overlap}

As a GPU schedules subgroup execution, data movement between global memory
and the GPU die may overlap with on-chip operations like arithmetic.  We
demonstrate how this nonlinear relationship between cost components and
overall cost can be accommodated in a model expression in the proposed
system.

If on-chip operations and global memory transactions overlap completely,
execution time may be approximated as
\begin{equation}
    t \approx \text{max}(c_\text{gmem}, c_\text{on-chip}),
    \label{eq:overlap_max}
\end{equation}
where $c_\text{gmem}$ and $c_\text{on-chip}$ are the time costs of global memory
transactions and on-chip operations, respectively. Since \perflex\ models must
be differentiable, we cannot use this approach directly. Instead, we use a
differentiable function $\hat{s}(x)$ approximating the step function
\begin{equation}
    s(x) =
    \begin{cases}
    0 & \text{if}\ x < 0, \\
    1 & \text{if}\ x \geq 0,
    \end{cases}
    \label{eq:step}
\end{equation}
to approximate the model described by \eqref{eq:overlap_max} as
\begin{align}
\begin{split}
    t \approx c_\text{gmem} \cdot &\hat{s}(c_\text{gmem}-c_\text{on-chip})\\
    +c_\text{on-chip} \cdot &\hat{s}(c_\text{on-chip}-c_\text{gmem}).
    \label{eq:overlap_differentiable}
\end{split}
\end{align}
Here, $\hat{s}(x)$ serves as a switch, ``turning on or off'' the global
memory and on-chip cost components, depending on which is greater. If
${c_\text{gmem} > c_\text{on-chip}}$, then
${\hat{s}(c_\text{gmem}-c_\text{on-chip}) \approx 1}$ and
${\hat{s}(c_\text{on-chip}-c_\text{gmem}) \approx 0}$, yielding ${t \approx
c_\text{gmem}}$. Alternatively, if on-chip costs dominate, ${t \approx
c_\text{on-chip}}$.

\begin{figure}
    \centering
    \includegraphics[width=0.345\textwidth]{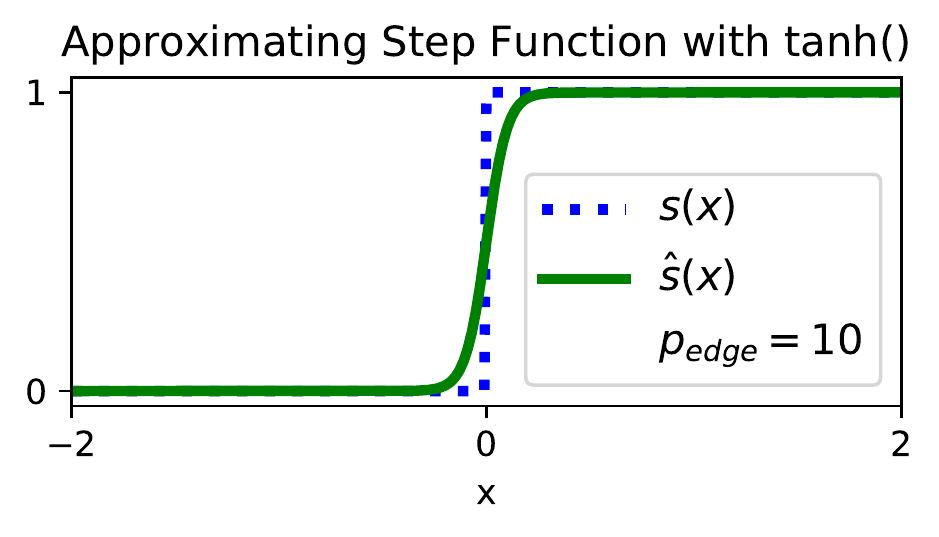}
    \caption{Approximating $s(x)$ with differentiable $\hat{s}(x)$.}
    \label{fig:tanh}
\end{figure}

To approximate the step function, we use
\begin{equation}
    \hat{s}(x) = (\text{tanh}(p_\text{edge} \cdot x)+1)/2,
    \label{eq:tanh}
\end{equation}
where $p_{edge}$ is a parameter determining the ``abruptness'' of the step.
It is determined along with the other parameters during model calibration. As
$p_\text{edge}$ increases, $\hat{s}(x)$ becomes more similar to $s(x)$.
Variations of \eqref{eq:tanh} with additional parameters could
model \emph{partial} overlap between global memory transactions and on-chip
operations as well. Figure~\ref{fig:tanh} shows a comparison between $s(x)$
and $\hat{s}(x)$ with ${p_\text{edge} = 10}$.

To demonstrate the effectiveness of this technique, we create a measurement
kernel in which we can vary the ratio of local to global memory accesses. In
this kernel, each thread performs one 32-bit global load, followed by $m$ 32-bit
local memory load-store sequences, followed by one 32-bit global store. By
varying $m$, the ratio of local to global memory accesses, we control whether
the execution time is dominated by global or local memory transactions.

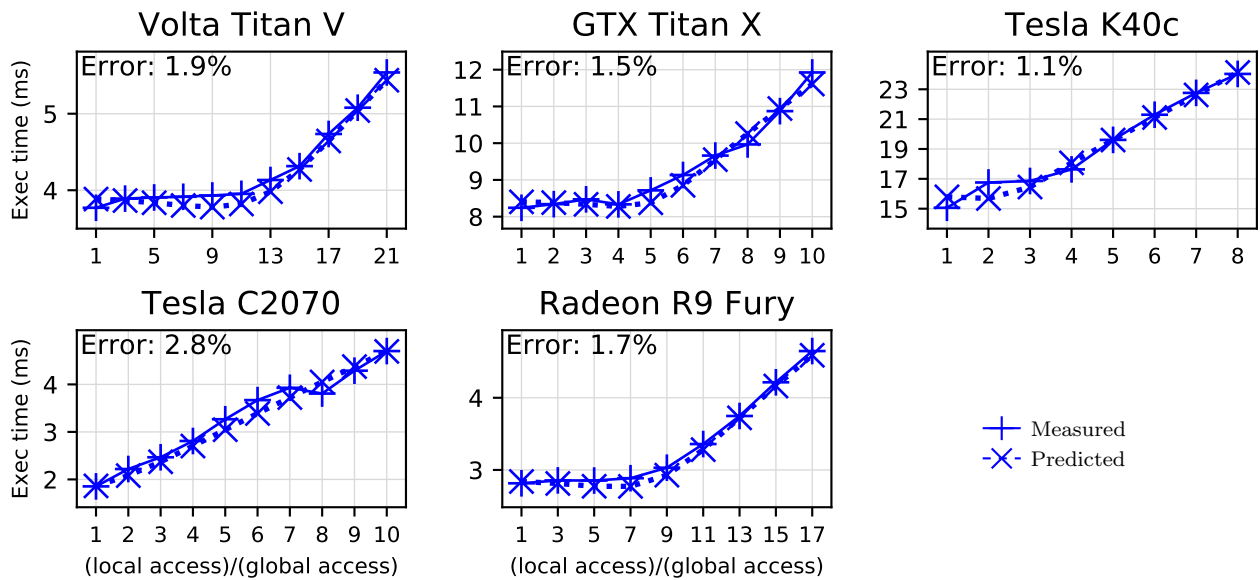
\begin{figure*}
\centering
\begin{subfigure}{.345\textwidth}
    \centering
    \includegraphics[trim={.25cm 0 .25cm .25cm},clip,right]{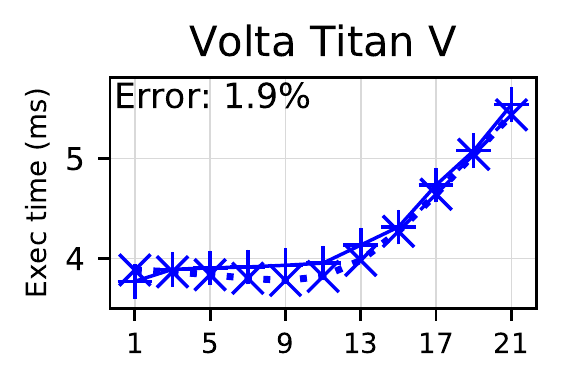}
\end{subfigure}
\begin{subfigure}{.315\textwidth}
    \centering
    \includegraphics[trim={.25cm 0 .25cm .25cm},clip,right]{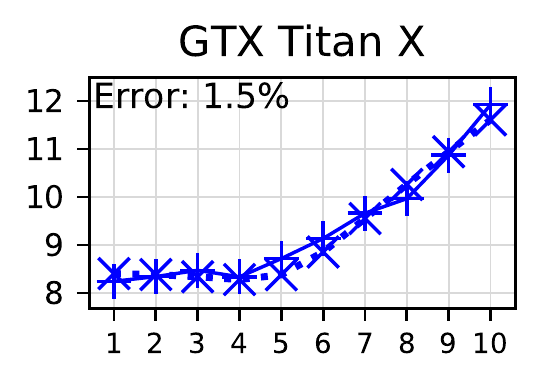}
\end{subfigure}
\begin{subfigure}{.315\textwidth}
    \centering
    \includegraphics[trim={.25cm 0 .25cm .25cm},clip,right]{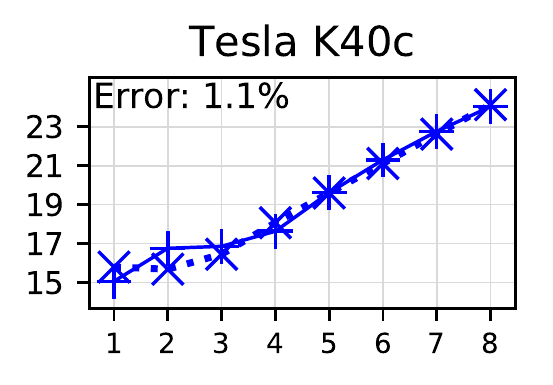}
\end{subfigure}
\begin{subfigure}{.345\textwidth}
    \centering
    \includegraphics[trim={.25cm 0 .25cm .25cm},clip,right]{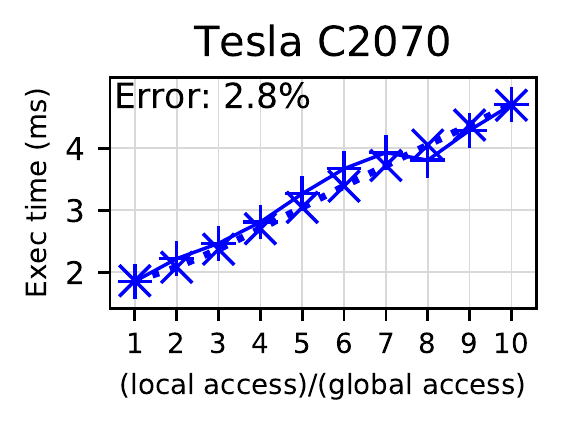}
\end{subfigure}
\begin{subfigure}{.315\textwidth}
    \centering
    \includegraphics[trim={.25cm 0 .25cm .25cm},clip,right]{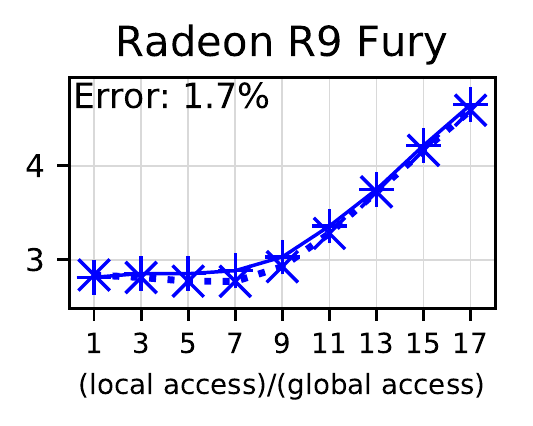}
\end{subfigure}
\begin{subfigure}{.315\textwidth}
    \centering
    \begin{tikzpicture}
      [
        nodestyle/.style={font=\footnotesize,thick}
      ]
        \node [nodestyle, anchor=west] (Measured) at (0.5, 0) {Measured};
        \draw [draw, thick, plotblue] (0,0) -- (Measured.west);
        \node [nodestyle, anchor=west] (Predicted) at (.5, -.4) {Predicted};
        \draw [draw, thick, dashed, plotblue] (0,-.4) -- (Predicted.west);
        \draw (0.25,0) node[cross=0.14cm, plotblue, rotate=45] {};
        \draw (0.25,-.4) node[cross=0.16cm, plotblue] {};
    \end{tikzpicture}
\end{subfigure}
\caption{Modeling overlap of local and global memory transactions. Geometric mean of relative error displayed. Array size differs across GPUs.}
\label{fig:local_global_overlap_kernel}
\end{figure*}

When $m$ is small, on-chip costs may be hidden behind global memory
transactions; as $m$ increases, eventually local memory transactions dominate
execution time. We model this using a \perflex\ model based on
\eqref{eq:overlap_differentiable}, with $c_\text{gmem}$ and
$c_\text{on-chip}$ being expressions containing \perflex\ features and
parameters representing the global and local memory access costs.
Figure~\ref{fig:local_global_overlap_kernel} displays how \perflex\ calibrates
such a model based on this data. We observe that the extent to which local
memory transaction costs in this kernel are hidden behind global transaction
costs varies significantly across machines. On the \nvidiakepler\ and
\nvidiafermi\ GPUs, very little, if any, of the local access cost is hidden,
while on the \nvidiavolta, \nvidiamaxwell, and \amdradeon\ GPUs, the cost of
anywhere from 4 to 12 local memory accesses can be hidden behind a global
transaction. We conclude that, at least for this kernel, on-chip and global
memory operation overlap behavior varies across GPUs, and that a \perflex\
model based on \eqref{eq:overlap_differentiable} can model this behavior. We
will discuss results for another kernel variant where this overlap behavior
varies across GPUs in Section~\ref{sec:dg_results}. OpenCL version and other
platform information is available in Table~\ref{table:gpus}.

\section{Results}
\label{sec:results}


To demonstrate our approach, we create and calibrate models for three
applications, with each model designed to predict execution times for
multiple variants of a particular computation. These computations include a
matrix-matrix multiplication (two variants), a Discontinuous Galerkin (DG)
differentiation operation (four variants), and a 2-D five-point finite
difference stencil computation (two variants). Using \uipick, we produce a
cost-analytic measurement kernel set for each model. This process decomposes
the computational cost incurred into individual cost components with parameters
calibrated by microbenchmarks which combine to model and thereby explain the
computational cost of the kernel under investigation. By conducting
measurements of our analytical microbenchmarks on the five GPUs in
Table~\ref{table:gpus}, we obtain calibrated models on each platform which we
then evaluate for predictiveness and accuracy. 

To obtain average execution times, we evaluate our OpenCL wall time
feature, discussed in Section~\ref{sec:other_features}. On the \amdradeon\
GPU, we observed that anomalous execution times on the order of 10$\times$
higher than a variant's usual execution time can occur occasionally, seemingly
at random, and we exclude these events from our data.

\begin{table}
    \centering
    \resizebox{\linewidth}{!}{%
    \begin{tabular}{ll}
        \toprule
        \textbf{GPU (Generation)} & \textbf{OpenCL/Platform/Driver Info}\\
        \midrule
        \nvidiavolta\ (Volta) & OCL 1.2, CUDA 10.0.246 (410.93)\\
        \nvidiamaxwell\ (Maxwell) & OCL 1.2, CUDA 10.0.292 (410.104)\\
        \nvidiakepler\ (Kepler) & OCL 1.2, CUDA 9.1.84 (390.87)\\
        \nvidiafermi\ (Fermi) & OCL 1.2 CUDA 9.1.84 (390.116)\\
        \amdradeon\ (GCN 3) & OpenCL/ROCm 1.2.0-2019020110\\
                            & ROCm platform HSA runtime 1.1.9-49-g39f1af5\\
                            & Kernel 4.19\\
        \bottomrule
    \end{tabular}
    }
    \caption{Platforms used for evaluation in Sections~\ref{sec:modeling_overlap} and \ref{sec:results}.}
    \label{table:gpus}
\end{table}

We compare execution times predicted by the models with measured execution
times in Figures~\ref{fig:mm_results}, \ref{fig:dg_results}, and
\ref{fig:fd_results}, and report the geometric mean of relative error for
reasons laid out by \cite{fleming_how_1986}.

\subsection{Models Demonstrated}
\label{sec:results_models}

In the models we consider in this evaluation, we categorize workload costs
as
\begin{itemize}
    \item $c_\text{gmem}$: global memory access,
    \item $c_\text{on-chip}$: on-chip work, i.e., local/scratchpad memory access
        and arithmetic, and
    \item $c_\text{overhead}$: barrier, kernel launch, and work-group launch costs.
\end{itemize}
\noindent
We model each of these three cost components individually as a sum of
kernel features weighted by cost parameters, with barrier cost modeled on a
per-work-group basis via the strategy described in
Section~\ref{sec:sync_features}. We evaluate two different types of models:
a \emph{linear} model,
\begin{equation}
    t \approx c_\text{overhead} + c_\text{gmem} + c_\text{on-chip}\text{,}
    \label{eq:linear_model}
\end{equation}
and a \emph{nonlinear} model that allows overlap of on-chip and global memory
operation costs,
\begin{align}
\begin{split}
    t \approx\ &c_\text{overhead}\ + \\
        &c_\text{gmem} \cdot \hat{s}(c_\text{gmem}-c_\text{on-chip})\ + \\
        &c_\text{on-chip} \cdot \hat{s}(c_\text{on-chip}-c_\text{gmem})\text{.}
    \label{eq:nonlinear_model}
\end{split}
\end{align}

In general, the extent to which on-chip operation costs are hidden by
global memory transactions varies between kernels and across architectures.
To determine whether the extent of this overlap warrants the nonlinear
model expressed in \eqref{eq:nonlinear_model}, we apply multiple
strategies.

First, before building and calibrating a performance model, we
use the work removal routine discussed in Section~\ref{sec:work_remover} to
remove arithmetic and local memory accesses from a kernel, obtaining
execution times for a version of the kernel containing only global memory
traffic. We then estimate the cost of the removed on-chip operations using
the costs revealed by our microbenchmark kernels. If the sum of these two
separate costs is approximately equal to the total execution time for the
original kernel, this suggests little to no overlap. However, if the sum of
these separate costs is significantly greater than the total execution
time, this serves as evidence that on-chip costs are being hidden.

To gain further confidence in the presence or absence of this overlap, we
can build and calibrate both kinds of execution time models and observe the
results. When we use the linear model to predict execution times for a
kernel exhibiting overlap of on-chip costs that are significant relative to
the total kernel execution time, we observe inflated predictions of
execution time, sometimes by a significant factor, as will be discussed in
Section~\ref{sec:mm_results}. We observe the opposite result when applying
the nonlinear model to a kernel where the cost of on-chip work is large
relative to total execution time and very little of this cost is hidden.
The development of an a-priori criterion that captures the extent of
overlap would streamline model selection and improve the predictiveness of
our evaluation models. We leave this for future work.

As mentioned in Section~\ref{sec:computing_params}, the least squares
residual can also serve as an indicator of model appropriateness, aiding in
the selection of a linear or nonlinear model.

\subsection{Measurement Kernel Sets for Evaluation}
\label{sec:results_mkernels}

The measurement kernel sets used to calibrate our models, unlike the set used
in the simple example in
Section~\ref{sec:model_creation_calibration_and_evaluation}, employ a
microbenchmarking approach and do \emph{not} include the computation
whose execution times we are predicting. Each microbenchmark kernel is
designed to reveal the cost associated with a single kernel feature. The
design of these measurement kernels was outlined in
Section~\ref{sec:mkernel_design}. We use the following notation when
referring to features:

\begin{equation*}
f\text{-\textcolor{red}{mem/op type}}^{\{\texttt{\textcolor{blue}{lid}}\ \text{\textcolor{blue}{strides}}\}\{\texttt{\textcolor{cyan}{gid}}\ \text{\textcolor{cyan}{strides}}\}}_{<\text{\textcolor{green}{data type}}>[\text{\textcolor{magenta}{AFR}}](\text{\textcolor{orange}{memory access tag}})}
\end{equation*}

To denote a measurement kernel exercising a particular feature, we substitute
the prefix \emph{k-} for the prefix \emph{f-}. The \emph{mem/op type} field
categorizes the feature as a global or local memory access, arithmetic
operation, local barrier, kernel launch, or work-group launch. For data motion
features, the \emph{stride}, \emph{data type}, and \emph{AFR} fields describe
the access pattern characteristics introduced in
Section~\ref{sec:data_motion_features}. The \emph{memory access tag} field
describes a memory access specified by a memory access tag as described in
Sections~\ref{sec:data_motion_features} and \ref{sec:mkernel_design}. 

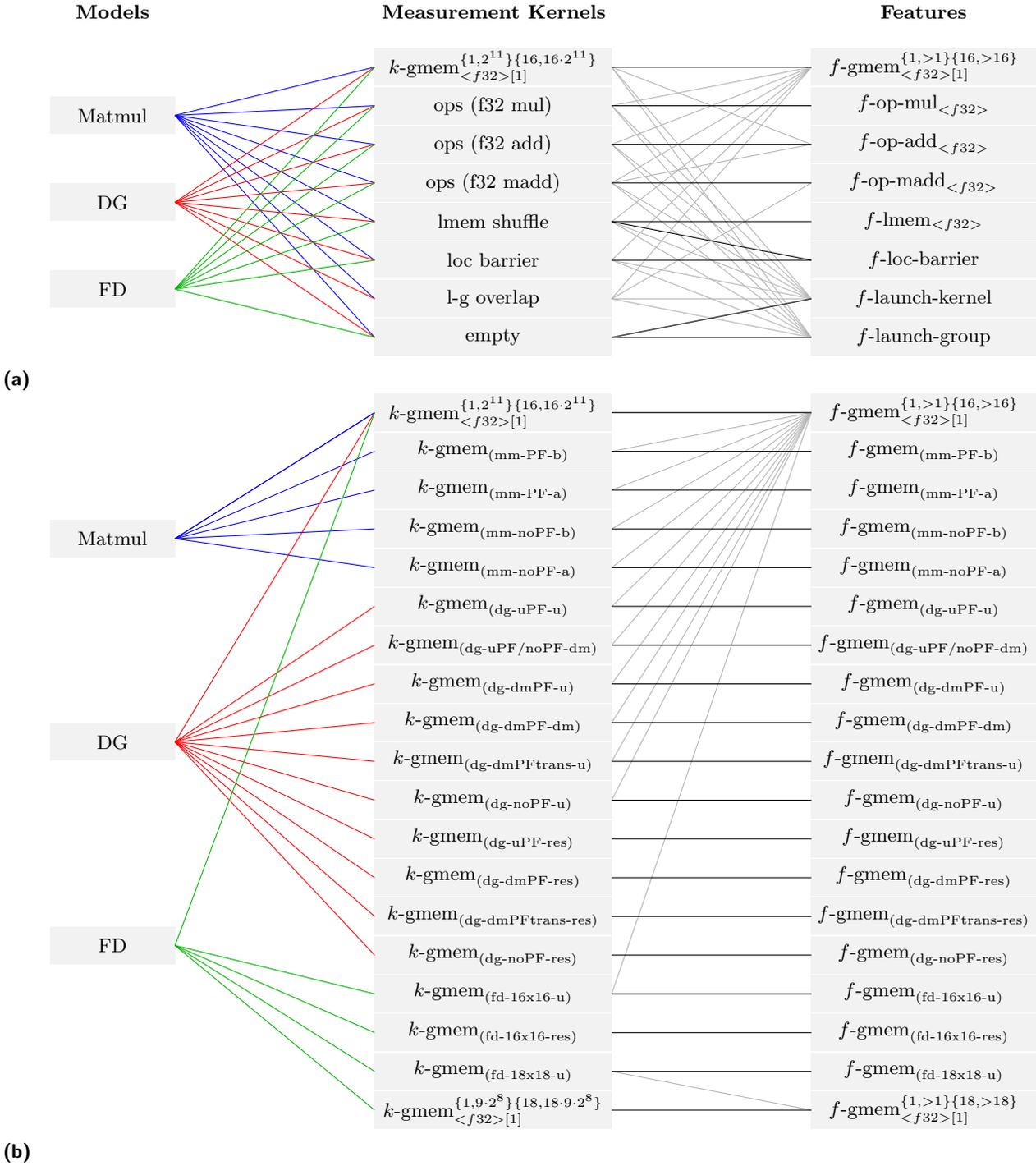
\begin{figure*}
    \begin{subfigure}{\textwidth}
    \centering
    \begin{tikzpicture}
[
  modelstyle/.style={line width=0, text=black, outer sep=0, font=\small, inner sep=0pt, minimum height=0.6cm, fill=black!5, minimum width=2.000000cm},
  mkernelstyle/.style={line width=0, text=black, outer sep=0, font=\small, inner sep=0pt, minimum height=0.6cm, fill=black!5, minimum width=3.800000cm},
  featstyle/.style={line width=0, text=black, outer sep=0, font=\small, inner sep=0pt, minimum height=0.6cm, fill=black!5, minimum width=3.600000cm},
  headerstyle_models/.style={line width=0, text=black, outer sep=0, font=\small, minimum width=2.000000cm},
  headerstyle_knls/.style={line width=0, text=black, outer sep=0, font=\small, minimum width=3.800000cm},
  headerstyle_feats/.style={line width=0, text=black, outer sep=0, font=\small, minimum width=3.600000cm},
]
  \node [modelstyle, anchor=north west] (Matmul) at (0.000000, -1.602500) {Matmul};
  \node [modelstyle, anchor=north west] (DG) at (0.000000, -3.005000) {DG};
  \node [modelstyle, anchor=north west] (FD) at (0.000000, -4.407500) {FD};
  \node [mkernelstyle, anchor=north west] (kernelmemgmem12111616cdot211f321) at (5.200000, -0.823333) {\kernelmem{gmem}{1}{2^{11}}{16}{16\cdot2^{11}}{f32}{1}};
  \node [mkernelstyle, anchor=north west] (ops f32 mul) at (5.200000, -1.446667) {ops (f32 mul)};
  \node [mkernelstyle, anchor=north west] (ops f32 add) at (5.200000, -2.070000) {ops (f32 add)};
  \node [mkernelstyle, anchor=north west] (ops f32 madd) at (5.200000, -2.693333) {ops (f32 madd)};
  \node [mkernelstyle, anchor=north west] (lmem shuffle) at (5.200000, -3.316667) {lmem shuffle};
  \node [mkernelstyle, anchor=north west] (loc barrier) at (5.200000, -3.940000) {loc barrier};
  \node [mkernelstyle, anchor=north west] (l-g overlap) at (5.200000, -4.563333) {l-g overlap};
  \node [mkernelstyle, anchor=north west] (empty) at (5.200000, -5.186667) {empty};
  \node [featstyle, anchor=north west] (featurememgmem111616f321) at (12.200000, -0.823333) {\featuremem{gmem}{1}{>1}{16}{>16}{f32}{1}};
  \node [featstyle, anchor=north west] (featureopop-mulf32) at (12.200000, -1.446667) {\featureop{op-mul}{f32}};
  \node [featstyle, anchor=north west] (featureopop-addf32) at (12.200000, -2.070000) {\featureop{op-add}{f32}};
  \node [featstyle, anchor=north west] (featureopop-maddf32) at (12.200000, -2.693333) {\featureop{op-madd}{f32}};
  \node [featstyle, anchor=north west] (featureoplmemf32) at (12.200000, -3.316667) {\featureop{lmem}{f32}};
  \node [featstyle, anchor=north west] (featuresimploc-barrier) at (12.200000, -3.940000) {\featuresimp{loc-barrier}};
  \node [featstyle, anchor=north west] (featuresimplaunch-kernel) at (12.200000, -4.563333) {\featuresimp{launch-kernel}};
  \node [featstyle, anchor=north west] (featuresimplaunch-group) at (12.200000, -5.186667) {\featuresimp{launch-group}};
  \node [headerstyle_models, anchor=north west] (textbfModels) at (0.000000, 0.000000) {\textbf{Models}};
  \node [headerstyle_knls, anchor=north west] (textbfMeasurement Kernels) at (5.200000, 0.000000) {\textbf{Measurement Kernels}};
  \node [headerstyle_feats, anchor=north west] (textbfFeatures) at (12.200000, 0.000000) {\textbf{Features}};

  \draw[blue] (Matmul.east) -- (kernelmemgmem12111616cdot211f321.west);
  \draw[blue] (Matmul.east) -- (ops f32 mul.west);
  \draw[blue] (Matmul.east) -- (ops f32 add.west);
  \draw[blue] (Matmul.east) -- (ops f32 madd.west);
  \draw[blue] (Matmul.east) -- (lmem shuffle.west);
  \draw[blue] (Matmul.east) -- (loc barrier.west);
  \draw[blue] (Matmul.east) -- (l-g overlap.west);
  \draw[blue] (Matmul.east) -- (empty.west);
  \draw[red] (DG.east) -- (kernelmemgmem12111616cdot211f321.west);
  \draw[red] (DG.east) -- (ops f32 mul.west);
  \draw[red] (DG.east) -- (ops f32 add.west);
  \draw[red] (DG.east) -- (ops f32 madd.west);
  \draw[red] (DG.east) -- (lmem shuffle.west);
  \draw[red] (DG.east) -- (loc barrier.west);
  \draw[red] (DG.east) -- (l-g overlap.west);
  \draw[red] (DG.east) -- (empty.west);
  \draw[green] (FD.east) -- (kernelmemgmem12111616cdot211f321.west);
  \draw[green] (FD.east) -- (ops f32 mul.west);
  \draw[green] (FD.east) -- (ops f32 add.west);
  \draw[green] (FD.east) -- (ops f32 madd.west);
  \draw[green] (FD.east) -- (lmem shuffle.west);
  \draw[green] (FD.east) -- (loc barrier.west);
  \draw[green] (FD.east) -- (empty.west);
  \draw[black!30] (featuresimplaunch-kernel.west) -- (kernelmemgmem12111616cdot211f321.east);
  \draw[black!30] (featuresimplaunch-kernel.west) -- (ops f32 mul.east);
  \draw[black!30] (featuresimplaunch-kernel.west) -- (ops f32 add.east);
  \draw[black!30] (featuresimplaunch-kernel.west) -- (ops f32 madd.east);
  \draw[black!30] (featuresimplaunch-kernel.west) -- (lmem shuffle.east);
  \draw[black!30] (featuresimplaunch-kernel.west) -- (loc barrier.east);
  \draw[black!30] (featuresimplaunch-kernel.west) -- (l-g overlap.east);
  \draw[black!30] (featuresimplaunch-group.west) -- (kernelmemgmem12111616cdot211f321.east);
  \draw[black!30] (featuresimplaunch-group.west) -- (ops f32 mul.east);
  \draw[black!30] (featuresimplaunch-group.west) -- (ops f32 add.east);
  \draw[black!30] (featuresimplaunch-group.west) -- (ops f32 madd.east);
  \draw[black!30] (featuresimplaunch-group.west) -- (lmem shuffle.east);
  \draw[black!30] (featuresimplaunch-group.west) -- (loc barrier.east);
  \draw[black!30] (featuresimplaunch-group.west) -- (l-g overlap.east);
  \draw[black!30] (l-g overlap.east) -- (featureopop-maddf32.west);
  \draw[black!30] (featureopop-addf32.west) -- (kernelmemgmem12111616cdot211f321.east);
  \draw[black!30] (featureopop-addf32.west) -- (ops f32 add.east);
  \draw[black!30] (featureopop-addf32.west) -- (ops f32 madd.east);
  \draw (empty.east) -- (featuresimplaunch-group.west);
  \draw (empty.east) -- (featuresimplaunch-kernel.west);
  \draw[black!30] (featurememgmem111616f321.west) -- (ops f32 mul.east);
  \draw[black!30] (featurememgmem111616f321.west) -- (ops f32 add.east);
  \draw[black!30] (featurememgmem111616f321.west) -- (ops f32 madd.east);
  \draw[black!30] (featurememgmem111616f321.west) -- (lmem shuffle.east);
  \draw[black!30] (featurememgmem111616f321.west) -- (loc barrier.east);
  \draw[black!30] (featurememgmem111616f321.west) -- (l-g overlap.east);
  \draw (featurememgmem111616f321.west) -- (kernelmemgmem12111616cdot211f321.east);
  \draw (ops f32 mul.east) -- (featureopop-mulf32.west);
  \draw (ops f32 add.east) -- (featureopop-addf32.west);
  \draw (ops f32 madd.east) -- (featureopop-maddf32.west);
  \draw (lmem shuffle.east) -- (featureoplmemf32.west);
  \draw (lmem shuffle.east) -- (featuresimploc-barrier.west);
  \draw (loc barrier.east) -- (featuresimploc-barrier.west);
\end{tikzpicture}
    \caption{}
    \label{fig:mkernels_common}
    \end{subfigure}
    \begin{subfigure}{\textwidth}
    \centering
    \begin{tikzpicture}
[
  modelstyle/.style={line width=0, text=black, outer sep=0, font=\small, inner sep=0pt, minimum height=0.6cm, fill=black!5, minimum width=2.000000cm},
  mkernelstyle/.style={line width=0, text=black, outer sep=0, font=\small, inner sep=0pt, minimum height=0.6cm, fill=black!5, minimum width=3.800000cm},
  featstyle/.style={line width=0, text=black, outer sep=0, font=\small, inner sep=0pt, minimum height=0.6cm, fill=black!5, minimum width=3.600000cm},
]
  \node [modelstyle, anchor=north west] (Matmul) at (0.000000, -2.685000) {Matmul};
  \node [modelstyle, anchor=north west] (DG) at (0.000000, -5.970000) {DG};
  \node [modelstyle, anchor=north west] (FD) at (0.000000, -9.255000) {FD};
  \node [mkernelstyle, anchor=north west] (kernelmemgmem12111616cdot211f321) at (5.200000, -0.651429) {\kernelmem{gmem}{1}{2^{11}}{16}{16\cdot2^{11}}{f32}{1}};
  \node [mkernelstyle, anchor=north west] (kernelmemtaggmemmm-PF-b) at (5.200000, -1.277143) {\kernelmemtag{gmem}{mm-PF-b}};
  \node [mkernelstyle, anchor=north west] (kernelmemtaggmemmm-PF-a) at (5.200000, -1.902857) {\kernelmemtag{gmem}{mm-PF-a}};
  \node [mkernelstyle, anchor=north west] (kernelmemtaggmemmm-noPF-b) at (5.200000, -2.528571) {\kernelmemtag{gmem}{mm-noPF-b}};
  \node [mkernelstyle, anchor=north west] (kernelmemtaggmemmm-noPF-a) at (5.200000, -3.154286) {\kernelmemtag{gmem}{mm-noPF-a}};
  \node [mkernelstyle, anchor=north west] (kernelmemtaggmemdg-uPF-u) at (5.200000, -3.780000) {\kernelmemtag{gmem}{dg-uPF-u}};
  \node [mkernelstyle, anchor=north west] (kernelmemtaggmemdg-uPF/noPF-dm) at (5.200000, -4.405714) {\kernelmemtag{gmem}{dg-uPF/noPF-dm}};
  \node [mkernelstyle, anchor=north west] (kernelmemtaggmemdg-dmPF-u) at (5.200000, -5.031429) {\kernelmemtag{gmem}{dg-dmPF-u}};
  \node [mkernelstyle, anchor=north west] (kernelmemtaggmemdg-dmPF-dm) at (5.200000, -5.657143) {\kernelmemtag{gmem}{dg-dmPF-dm}};
  \node [mkernelstyle, anchor=north west] (kernelmemtaggmemdg-dmPFtrans-u) at (5.200000, -6.282857) {\kernelmemtag{gmem}{dg-dmPFtrans-u}};
  \node [mkernelstyle, anchor=north west] (kernelmemtaggmemdg-noPF-u) at (5.200000, -6.908571) {\kernelmemtag{gmem}{dg-noPF-u}};
  \node [mkernelstyle, anchor=north west] (kernelmemtaggmemdg-uPF-res) at (5.200000, -7.534286) {\kernelmemtag{gmem}{dg-uPF-res}};
  \node [mkernelstyle, anchor=north west] (kernelmemtaggmemdg-dmPF-res) at (5.200000, -8.160000) {\kernelmemtag{gmem}{dg-dmPF-res}};
  \node [mkernelstyle, anchor=north west] (kernelmemtaggmemdg-dmPFtrans-res) at (5.200000, -8.785714) {\kernelmemtag{gmem}{dg-dmPFtrans-res}};
  \node [mkernelstyle, anchor=north west] (kernelmemtaggmemdg-noPF-res) at (5.200000, -9.411429) {\kernelmemtag{gmem}{dg-noPF-res}};
  \node [mkernelstyle, anchor=north west] (kernelmemtaggmemfd-16x16-u) at (5.200000, -10.037143) {\kernelmemtag{gmem}{fd-16x16-u}};
  \node [mkernelstyle, anchor=north west] (kernelmemtaggmemfd-16x16-res) at (5.200000, -10.662857) {\kernelmemtag{gmem}{fd-16x16-res}};
  \node [mkernelstyle, anchor=north west] (kernelmemtaggmemfd-18x18-u) at (5.200000, -11.288571) {\kernelmemtag{gmem}{fd-18x18-u}};
  \node [mkernelstyle, anchor=north west] (kernelmemgmem19cdot281818cdot9cdot28f321) at (5.200000, -11.914286) {\kernelmem{gmem}{1}{9\cdot2^8}{18}{18\cdot9\cdot2^8}{f32}{1}};
  \node [featstyle, anchor=north west] (featurememgmem111616f321) at (12.200000, -0.651429) {\featuremem{gmem}{1}{>1}{16}{>16}{f32}{1}};
  \node [featstyle, anchor=north west] (featurememtaggmemmm-PF-b) at (12.200000, -1.277143) {\featurememtag{gmem}{mm-PF-b}};
  \node [featstyle, anchor=north west] (featurememtaggmemmm-PF-a) at (12.200000, -1.902857) {\featurememtag{gmem}{mm-PF-a}};
  \node [featstyle, anchor=north west] (featurememtaggmemmm-noPF-b) at (12.200000, -2.528571) {\featurememtag{gmem}{mm-noPF-b}};
  \node [featstyle, anchor=north west] (featurememtaggmemmm-noPF-a) at (12.200000, -3.154286) {\featurememtag{gmem}{mm-noPF-a}};
  \node [featstyle, anchor=north west] (featurememtaggmemdg-uPF-u) at (12.200000, -3.780000) {\featurememtag{gmem}{dg-uPF-u}};
  \node [featstyle, anchor=north west] (featurememtaggmemdg-uPF/noPF-dm) at (12.200000, -4.405714) {\featurememtag{gmem}{dg-uPF/noPF-dm}};
  \node [featstyle, anchor=north west] (featurememtaggmemdg-dmPF-u) at (12.200000, -5.031429) {\featurememtag{gmem}{dg-dmPF-u}};
  \node [featstyle, anchor=north west] (featurememtaggmemdg-dmPF-dm) at (12.200000, -5.657143) {\featurememtag{gmem}{dg-dmPF-dm}};
  \node [featstyle, anchor=north west] (featurememtaggmemdg-dmPFtrans-u) at (12.200000, -6.282857) {\featurememtag{gmem}{dg-dmPFtrans-u}};
  \node [featstyle, anchor=north west] (featurememtaggmemdg-noPF-u) at (12.200000, -6.908571) {\featurememtag{gmem}{dg-noPF-u}};
  \node [featstyle, anchor=north west] (featurememtaggmemdg-uPF-res) at (12.200000, -7.534286) {\featurememtag{gmem}{dg-uPF-res}};
  \node [featstyle, anchor=north west] (featurememtaggmemdg-dmPF-res) at (12.200000, -8.160000) {\featurememtag{gmem}{dg-dmPF-res}};
  \node [featstyle, anchor=north west] (featurememtaggmemdg-dmPFtrans-res) at (12.200000, -8.785714) {\featurememtag{gmem}{dg-dmPFtrans-res}};
  \node [featstyle, anchor=north west] (featurememtaggmemdg-noPF-res) at (12.200000, -9.411429) {\featurememtag{gmem}{dg-noPF-res}};
  \node [featstyle, anchor=north west] (featurememtaggmemfd-16x16-u) at (12.200000, -10.037143) {\featurememtag{gmem}{fd-16x16-u}};
  \node [featstyle, anchor=north west] (featurememtaggmemfd-16x16-res) at (12.200000, -10.662857) {\featurememtag{gmem}{fd-16x16-res}};
  \node [featstyle, anchor=north west] (featurememtaggmemfd-18x18-u) at (12.200000, -11.288571) {\featurememtag{gmem}{fd-18x18-u}};
  \node [featstyle, anchor=north west] (featurememgmem111818f321) at (12.200000, -11.914286) {\featuremem{gmem}{1}{>1}{18}{>18}{f32}{1}};

  \draw[blue] (Matmul.east) -- (kernelmemgmem12111616cdot211f321.west);
  \draw[blue] (Matmul.east) -- (kernelmemtaggmemmm-PF-b.west);
  \draw[blue] (Matmul.east) -- (kernelmemtaggmemmm-PF-a.west);
  \draw[blue] (Matmul.east) -- (kernelmemtaggmemmm-noPF-b.west);
  \draw[blue] (Matmul.east) -- (kernelmemtaggmemmm-noPF-a.west);
  \draw[blue] (Matmul.east) -- (kernelmemgmem12111616cdot211f321.west);
  \draw[red] (DG.east) -- (kernelmemtaggmemdg-uPF-u.west);
  \draw[red] (DG.east) -- (kernelmemtaggmemdg-uPF/noPF-dm.west);
  \draw[red] (DG.east) -- (kernelmemtaggmemdg-uPF-res.west);
  \draw[red] (DG.east) -- (kernelmemtaggmemdg-dmPF-u.west);
  \draw[red] (DG.east) -- (kernelmemtaggmemdg-dmPF-dm.west);
  \draw[red] (DG.east) -- (kernelmemtaggmemdg-dmPF-res.west);
  \draw[red] (DG.east) -- (kernelmemtaggmemdg-dmPFtrans-u.west);
  \draw[red] (DG.east) -- (kernelmemtaggmemdg-dmPFtrans-res.west);
  \draw[red] (DG.east) -- (kernelmemtaggmemdg-noPF-u.west);
  \draw[red] (DG.east) -- (kernelmemtaggmemdg-noPF-res.west);
  \draw[red] (DG.east) -- (kernelmemgmem12111616cdot211f321.west);
  \draw[green] (FD.east) -- (kernelmemtaggmemfd-16x16-u.west);
  \draw[green] (FD.east) -- (kernelmemtaggmemfd-16x16-res.west);
  \draw[green] (FD.east) -- (kernelmemtaggmemfd-18x18-u.west);
  \draw[green] (FD.east) -- (kernelmemgmem19cdot281818cdot9cdot28f321.west);
  \draw[green] (FD.east) -- (kernelmemgmem12111616cdot211f321.west);
  \draw[black!30] (featurememgmem111616f321.west) -- (kernelmemgmem12111616cdot211f321.east);
  \draw[black!30] (featurememgmem111616f321.west) -- (kernelmemtaggmemmm-PF-b.east);
  \draw[black!30] (featurememgmem111616f321.west) -- (kernelmemtaggmemmm-PF-a.east);
  \draw[black!30] (featurememgmem111616f321.west) -- (kernelmemtaggmemmm-noPF-b.east);
  \draw[black!30] (featurememgmem111616f321.west) -- (kernelmemtaggmemmm-noPF-a.east);
  \draw[black!30] (featurememgmem111616f321.west) -- (kernelmemtaggmemdg-uPF-u.east);
  \draw[black!30] (featurememgmem111616f321.west) -- (kernelmemtaggmemdg-uPF/noPF-dm.east);
  \draw[black!30] (featurememgmem111616f321.west) -- (kernelmemtaggmemdg-dmPF-u.east);
  \draw[black!30] (featurememgmem111616f321.west) -- (kernelmemtaggmemdg-dmPF-dm.east);
  \draw[black!30] (featurememgmem111616f321.west) -- (kernelmemtaggmemdg-dmPFtrans-u.east);
  \draw[black!30] (featurememgmem111616f321.west) -- (kernelmemtaggmemdg-noPF-u.east);
  \draw[black!30] (featurememgmem111616f321.west) -- (kernelmemtaggmemfd-16x16-u.east);
  \draw[black!30] (kernelmemtaggmemfd-18x18-u.east) -- (featurememgmem111818f321.west);
  \draw (kernelmemgmem12111616cdot211f321.east) -- (featurememgmem111616f321.west);
  \draw (kernelmemtaggmemmm-PF-b.east) -- (featurememtaggmemmm-PF-b.west);
  \draw (kernelmemtaggmemmm-PF-a.east) -- (featurememtaggmemmm-PF-a.west);
  \draw (kernelmemtaggmemmm-noPF-b.east) -- (featurememtaggmemmm-noPF-b.west);
  \draw (kernelmemtaggmemmm-noPF-a.east) -- (featurememtaggmemmm-noPF-a.west);
  \draw (kernelmemtaggmemdg-uPF-u.east) -- (featurememtaggmemdg-uPF-u.west);
  \draw (kernelmemtaggmemdg-uPF/noPF-dm.east) -- (featurememtaggmemdg-uPF/noPF-dm.west);
  \draw (kernelmemtaggmemdg-uPF-res.east) -- (featurememtaggmemdg-uPF-res.west);
  \draw (kernelmemtaggmemdg-dmPF-u.east) -- (featurememtaggmemdg-dmPF-u.west);
  \draw (kernelmemtaggmemdg-dmPF-dm.east) -- (featurememtaggmemdg-dmPF-dm.west);
  \draw (kernelmemtaggmemdg-dmPF-res.east) -- (featurememtaggmemdg-dmPF-res.west);
  \draw (kernelmemtaggmemdg-dmPFtrans-u.east) -- (featurememtaggmemdg-dmPFtrans-u.west);
  \draw (kernelmemtaggmemdg-dmPFtrans-res.east) -- (featurememtaggmemdg-dmPFtrans-res.west);
  \draw (kernelmemtaggmemdg-noPF-u.east) -- (featurememtaggmemdg-noPF-u.west);
  \draw (kernelmemtaggmemdg-noPF-res.east) -- (featurememtaggmemdg-noPF-res.west);
  \draw (kernelmemtaggmemfd-16x16-u.east) -- (featurememtaggmemfd-16x16-u.west);
  \draw (kernelmemtaggmemfd-16x16-res.east) -- (featurememtaggmemfd-16x16-res.west);
  \draw (kernelmemtaggmemfd-18x18-u.east) -- (featurememtaggmemfd-18x18-u.west);
  \draw (kernelmemgmem19cdot281818cdot9cdot28f321.east) -- (featurememgmem111818f321.west);
\end{tikzpicture}
    \caption{}
    \label{fig:mkernels_other}
    \end{subfigure}
    \caption{Measurement kernels used in the models used for evaluation.
        Lighter grey lines connect a measurement kernel to features present in
        the kernel that are not the primary feature being targeted for
        measurement. Figure~\ref{fig:mkernels_common} shows measurement kernels
        used to calibrate non-global-memory-access parameters.
        Figure~\ref{fig:mkernels_other} shows measurement kernels used to
        calibrate global memory access parameters. Not shown: all gmem
        measurement kernels also contain \featuresimp{launch-kernel},
\featuresimp{launch-group}, and \featureop{op-add}{f32} features.}
    \label{fig:mkernels}
\end{figure*}

Figure~\ref{fig:mkernels} shows which measurement kernels we use to calibrate
each model, as well as the features present in the model for each kernel.


\subsection{Matrix Multiplication}
\label{sec:mm_results}


Our first evaluation model predicts execution times for two variants of square
matrix-matrix multiplication. The first variant, which is often used as an
introductory example in teaching GPU programming, is the same as the one
discussed in Section~\ref{sec:kernel_creation_and_transformation}. It
prefetches tiles of the two input matrices into local (shared) memory before
performing arithmetic. The second is the same algorithm without any
prefetching, and without splitting of the \texttt{k} summation loop. The
prefetching variant achieves between 8\% and 20\% of peak FLOP/s rates on all
five GPUs.
It is important to clarify the relationship between these measurements and the
validity assumptions (in particular, on machine utilization) set forth in
Section~\ref{sec:assumptions}.
The utilization assumption does not entail that all kernels to
which our modeling approach is applicable must already achieve peak performance, as this
would not reflect the typical use case of performance modeling. Instead, the
assumption exists to highlight potential sources of model
inaccuracy or performance shortfall, as revealed by the model.

Both variants of our matrix-matrix multiplication evaluation case operate on
32-bit floating point data and use $16\times16$ work-groups. Together, the two
variants use five distinct global memory access patterns, as shown in
Figure~\ref{fig:mkernels_other}. We model execution time using the nonlinear
model expressed in \eqref{eq:nonlinear_model}.  The model features comprising
$c_\text{gmem}$, $c_\text{on-chip}$, and $c_\text{overhead}$, as well as the
measurement kernel set, are shown in Figure~\ref{fig:mkernels}.

\setlength{\tabcolsep}{0.13cm}
\begin{table}
\begin{center}
\begin{tabular}{lccr}
    \toprule
    & \multicolumn{1}{c}{\textbf{Param.}} & &\\
    \textbf{Feature} & \textbf{val. (s)} & \textbf{MCG} & \multicolumn{1}{c}{\textbf{Rate}}\\
    \midrule
    f32 add & $5.4\mathrm{e}{-12}$ & SG & $5.9\mathrm{e}{+12}$ op/s\\
    f32 mul & $5.4\mathrm{e}{-12}$ & SG & $5.9\mathrm{e}{+12}$ op/s\\
    f32 madd & $5.0\mathrm{e}{-12}$ & SG & $12.8\mathrm{e}{+12}$ op/s\\
    \small{\featureop{lmem}{f32}} & $9.5\mathrm{e}{-12}$ & SG & $1.3\mathrm{e}{+13}$ B/s\\
    \small{\featuremem{gmem}{1}{>1}{16}{>16}{f32}{1}} & $3.5\mathrm{e}{-12}$ & WI & $1.1\mathrm{e}{+12}$ B/s\\
    \small{\featurememtag{gmem}{mm-PF-b}} & $4.8\mathrm{e}{-12}$ & WI & $8.3\mathrm{e}{+11}$ B/s\\
    \small{\featurememtag{gmem}{mm-PF-a}} & $1.9\mathrm{e}{-12}$ & WI & $2.1\mathrm{e}{+12}$ B/s\\
    \small{\featurememtag{gmem}{mm-noPF-b}} & $8.6\mathrm{e}{-13}$ & WI & $4.7\mathrm{e}{+12}$ B/s\\
    \small{\featurememtag{gmem}{mm-noPF-a}} & $1.9\mathrm{e}{-11}$ & SG & $6.7\mathrm{e}{+12}$ B/s\\
    local barrier & $1.3\mathrm{e}{-13}$ & WG &
        \multirow{4}{*}{\fbox{\begin{varwidth}{\textwidth} \centering \small{\textbf{Peak}}\\\scriptsize{$12.3\mathrm{e}{+12}$ FLOP/s}\\\scriptsize{$6.5\mathrm{e}{+11}$ B/s} \end{varwidth}}}\\
    launch group & $1.6\mathrm{e}{-09}$ & WG &\\
    launch kernel & $7.7\mathrm{e}{-05}$ & K &\\
    ($p_\text{edge}$) & $1.3\mathrm{e}{+03}$ & N/A &\\ 
    \bottomrule
\end{tabular}
\caption{Matrix multiplication model parameter values on the \nvidiavolta\ GPU.
    Parameter values represent costs per unit according to granularities
    listed, with the exception of `overlap edge', which is the parameter
    governing the sharpness of our step function approximation, $p_\text{edge}$
    in Equation~\ref{eq:tanh}. Modeled cost granularity (MCG): work-item (WI),
    sub-group (SG), work-group (WG), or kernel (K). Peak values obtained from
    \cite{nvidiaWiki}.}
\label{table:mm_param_vals}
\end{center}
\end{table}
\setlength{\tabcolsep}{\defaulttabcolsep}

Table~\ref{table:mm_param_vals} displays the model parameter values
representing feature costs on the \nvidiavolta\ GPU, as well as $p_\text{edge}$
from \eqref{eq:tanh}, whose value is determined along with the feature
cost parameters during the calibration process. Recall that these values aim to
represent \emph{effective} costs at maximum throughput. The units of work whose
cost we model are determined by the modeled cost granularities (MCGs)
listed; each operation cost modeled is assessed per work-item (WI), sub-group
(SG), work-group (WG), or kernel (K). Note that we count
\featurememtag{gmem}{mm-noPF-a} once per sub-group (32 work-items) rather than
once per work-item because the \texttt{lid(0)} stride is 0, as discussed in
Section~\ref{sec:stats}. The table also displays a throughput calculated based
on each parameter value and corresponding MCG. 

In this example set of parameter values, we observe similar costs for addition,
multiplication, and multiply-add operations, as we expect, and that the local
memory access cost is about twice that of the arithmetic operations. We also
observe that the throughputs implied by the arithmetic operation
costs match the peak 32-bit FLOP/s rate for the hardware very closely. Note that
the peak hardware FLOP/s rate listed here assumes multiply-add operations are
counted as two operations.

The accessibility of these parameter values and the transparency of the costs
they represent can facilitate understanding of the factors affecting
performance in these kernels. For example, by comparing the data throughput 
for \emph{mm-PF-a} and \emph{mm-PF-b} to the throughputs for
\emph{mm-noPF-a} and \emph{mm-noPF-b}, we observe that prefetching
\emph{increases} the effective cost per item loaded from global memory by
factors of about 3 and 5, respectively. This suggests the overall cost savings
due to prefetching is primarily due to the reduction in total number of global
memory accesses by a factor of the tile width, 16, rather than by a reduction
in cost of individual memory accesses. 

Since the access-to-footprint ratios for matrices \texttt{a} and \texttt{b} are
significantly greater than one (\texttt{n}, for the non-prefetching case), it
is not unreasonable that the apparent data throughput is greater than the
hardware peak. Elements of these arrays may be reused rather than re-fetched
from global memory due to, e.g., hardware caching, inflating the calculated
throughput. In calibrating these models, we have further repeatedly made the
peculiar observation that memory access patterns with an AFR of 1, such as
\featuremem{gmem}{1}{>1}{16}{>16}{f32}{1} in Table~\ref{table:mm_param_vals},
attain calibration parameters corresponding to slightly higher-than-peak rates.
This is a potentially interesting phenomenon whose further examination we leave
for future work. 

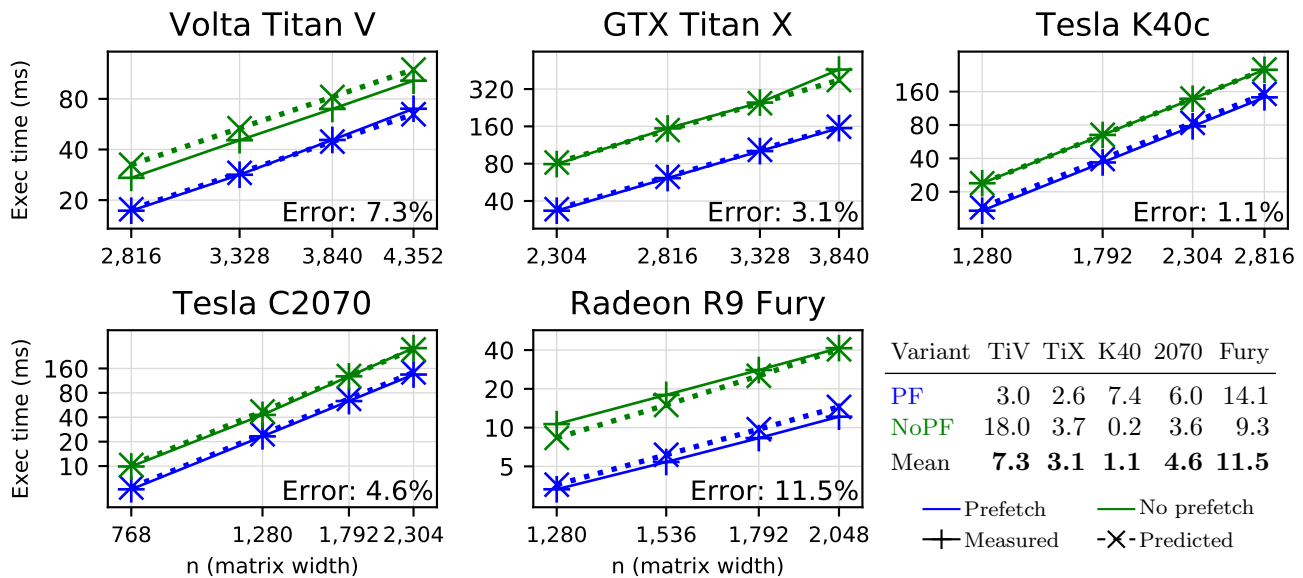
\begin{figure*}
\centering
\begin{subfigure}{.345\textwidth}
    \includegraphics[trim={.25cm 0 .25cm .25cm},clip,right]{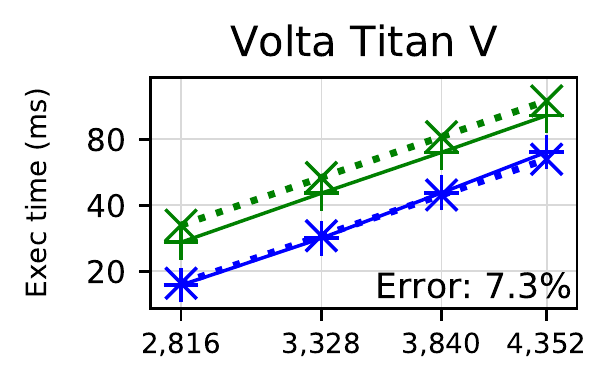}
\end{subfigure}
\begin{subfigure}{.315\textwidth}
    \includegraphics[trim={.25cm 0 .25cm .25cm},clip,right]{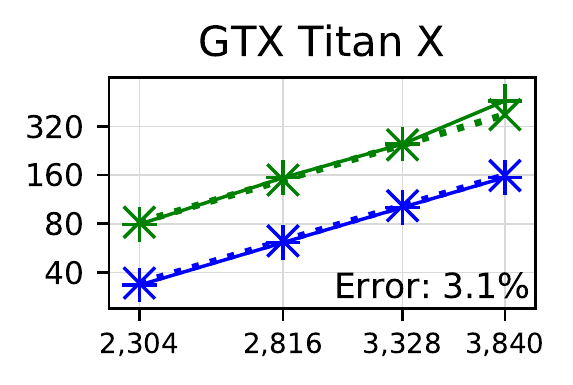}
\end{subfigure}
\begin{subfigure}{.315\textwidth}
    \includegraphics[trim={.25cm 0 .25cm .25cm},clip,right]{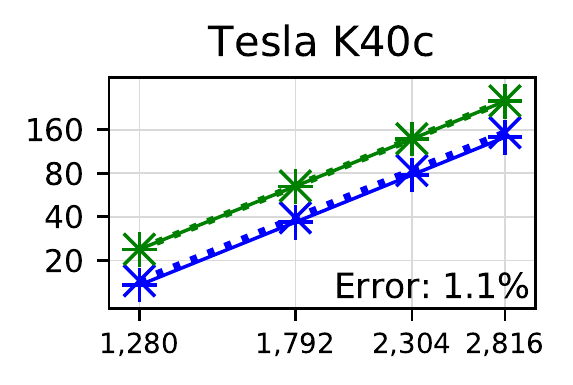}
\end{subfigure}
\begin{subfigure}{.345\textwidth}
    \includegraphics[trim={.25cm 0 .25cm .25cm},clip,right]{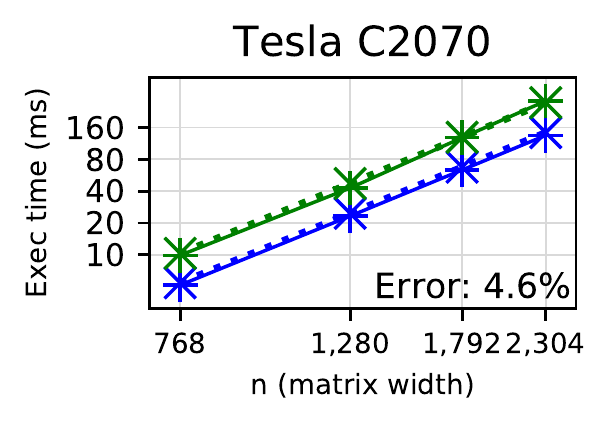}
\end{subfigure}
\begin{subfigure}{.315\textwidth}
    \includegraphics[trim={.25cm 0 .25cm .25cm},clip,right]{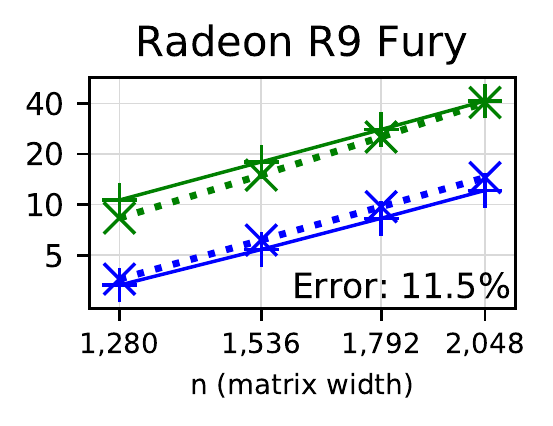}
\end{subfigure}
\begin{subfigure}{.315\textwidth}
    \centering
    \setlength{\tabcolsep}{0.08cm}
\resizebox{0.96\linewidth}{!}{
\begin{tabular}{lrrrrr}
  {\small Variant} & {\small TiV} & {\small TiX} & {\small K40} & {\small 2070} & {\small Fury} \\
  \midrule
  {\color{plotblue} \small PF} & 3.0 & 2.6 & 7.4 & 6.0 & 14.1 \\
  {\color{plotgreen} \small NoPF} & 18.0 & 3.7 & 0.2 & 3.6 & 9.3 \\
  \addlinespace[1pt]
  {\small Mean} & \textbf{7.3} & \textbf{3.1} & \textbf{1.1} & \textbf{4.6} & \textbf{11.5} \\
\end{tabular}
}
\setlength{\tabcolsep}{\defaulttabcolsep}
 \\
    \vspace{0.2cm}
    \begin{tikzpicture}
      [
        nodestyle/.style={font=\footnotesize,thick,inner sep=1pt}
      ]

        \node [nodestyle, anchor=west] (Prefetch) at (0.5, 0) {Prefetch};
        \draw [draw, thick, plotblue] (0, 0) -- (Prefetch.west);
        \node [nodestyle, anchor=west] (Noprefetch) at (2.8, 0) {No prefetch};
        \draw [draw, thick, plotgreen] (2.3, 0) -- (Noprefetch.west);

        \node [nodestyle, anchor=west] (Measured) at (0.5, -.4) {Measured};
        \draw [draw, thick] (0,-.4) -- (Measured.west);
        \node [nodestyle, anchor=west] (Predicted) at (2.8, -.4) {Predicted};
        \draw [draw, thick, dashed] (2.3,-.4) -- (Predicted.west);
        \draw (0.25,-.4) node[cross=0.12cm, rotate=45] {};
        \draw (2.55,-.4) node[cross=0.14cm] {};
    \end{tikzpicture}
\end{subfigure}
\caption{Matrix multiplication model accuracy. The table displays the geometric mean of relative error (\%).}
\label{fig:mm_results}
\end{figure*}

Figure~\ref{fig:mm_results} compares modeled to measured execution times for
the two variants on five GPUs, and displays the geometric mean of relative
error across both variants on each individual GPU as well as the error for
each individual variant-GPU combination across a range of problem sizes. On all
five GPUs, the model predicts the execution times of these variants accurately
enough to determine which variant is faster, with less than 10\% error in most
cases. Across all cases, the geometric mean of the relative error is 4.3\%.

For exploration, we also observed the predictions that would have been
made by the \emph{linear} model in \eqref{eq:linear_model}.
Using this model, the error for the non-prefetching variant is similar,
likely due to the on-chip costs being relatively small in comparison to the
total execution time. The linear model however over-predicts execution
time for the prefetching variant by between 40\% and 110\% on all GPUs. One way
to interpret this observation is that the prefetching variant, which performs
$2 \cdot n^3$ local memory loads and $2 \cdot n^3/16$ local memory stores that
are not performed in the non-prefetching variant, and 16 times \emph{fewer}
global memory loads, is successfully hiding the cost of local memory
transactions and arithmetic operations behind global memory transactions, and
that the nonlinear model is a good choice for this computation on these
architectures. Results of the on-chip-cost-hiding analysis described in
Section~\ref{sec:results_models} are consistent with this interpretation.

\subsection{DG Differentiation}
\label{sec:dg_results}

The discontinuous Galerkin (DG) finite element method
\citep{hesthaven2007nodal} is a numerical method for the approximate,
high-order accurate solution of boundary value problems of wave-like
(hyperbolic) PDEs (Partial Differential Equations) such as Euler's or Maxwell's
equations, often used in complex geometry on unstructured meshes.  Our second
evaluation case models execution times for four variants of an element-wise
differentiation of per-element polynomials used in a DG computation. Some of
these are derived from parallelization schemes in \cite{klockner2009nodal}. The
pre-transform \loopy\ kernel shows the mathematical operation performed by all
of these variants:
\begin{tcolorbox}[listingbox]
\begin{lstlisting}[style=custompython]
knl = lp.make_kernel(
    "{[m,k,i,j]: 0<=m<nmatrices and 0<=k<nelements and 0<=i,j<nunit_nodes}",
    "res[m,k,i] = sum(j, diff_mat[m,i,j] * u[k,j])")
\end{lstlisting}
\end{tcolorbox}


For a single differentiation matrix \texttt{diff\_mat} (i.e.,
\texttt{nmatrices} $=1$), this operation can be viewed as a matrix-matrix
multiplication where a small square matrix is multiplied by a `short and wide'
element matrix \texttt{u}. The primary differences between this computation and
that of the previous example are (1), typically, $\texttt{nunit\_nodes} <<
\texttt{nelements}$, yielding small \texttt{diff\_mat} matrices and a `short
and wide' \texttt{u}, and (2), since \texttt{nmatrices} $>1$, the operation on
\texttt{u} is performed multiple times, providing opportunities for data reuse.


All four variants tile and parallelize the \texttt{k} and \texttt{i} loops:
\begin{tcolorbox}[listingbox]
\begin{lstlisting}[style=custompython]
knl = lp.split_iname(knl,
    "i", lsize[1], outer_tag="g.1", inner_tag="l.1")
knl = lp.split_iname(knl,
    "k", lsize[0], outer_tag="g.0", inner_tag="l.0")
\end{lstlisting}
\end{tcolorbox}
\noindent
The first variant performs only these transformations and does not utilize
local memory for data reuse.

The second variant prefetches \texttt{lsize[0]} $\times$ \texttt{lsize[1]}
($16\times16$) tiles from \texttt{u} into local (shared) memory before
performing arithmetic:
\begin{tcolorbox}[listingbox]
\begin{lstlisting}[style=custompython]
... # (first split and tag i and k as above)
knl = lp.split_iname(knl, "j", lsize[1])
knl = lp.add_prefetch(knl, "u", ["k_in", "j_in"])
knl = lp.fix_parameters(knl, nmatrices=nmatrices)
knl = lp.add_inames_to_insn(knl, "i_out", "id:*fetch*")
knl = lp.realize_reduction(knl)
knl = lp.privatize_temporaries_with_inames(knl,"m")
knl = lp.duplicate_inames(knl, "m", "id:*init*")
knl = lp.duplicate_inames(knl, "m", "writes:res")
knl = lp.prioritize_loops(knl, ["j_out", "j_in", "m"])
\end{lstlisting}
\end{tcolorbox}
\noindent
These additional transformations tile the \texttt{j} loop, load parts of
\texttt{u} into scratchpad (local) memory, and restructure the loops to expose
instruction-level parallelism.

The third variant instead prefetches \texttt{lsize[0]} $\times$
\texttt{lsize[1]} tiles from the differentiation matrix \texttt{diff\_mat} into
local memory before performing arithmetic:
\begin{tcolorbox}[listingbox]
\begin{lstlisting}[style=custompython]
... # (first split and tag i and k as above)
knl = lp.split_iname(knl, "j", lsize[0])
knl = lp.add_prefetch(knl, "diff_mat", ["j_in","i_in"])
knl = lp.prioritize_loops(knl, ["m", "j_out", "j_in"])
\end{lstlisting}
\end{tcolorbox}
\noindent

The fourth variant uses the same transformations as the third; however, we
also transpose the memory layout of the element data:
\begin{tcolorbox}[listingbox]
\begin{lstlisting}[style=custompython]
knl = lp.tag_data_axes(knl, "u", "N0,N1")
knl = lp.tag_data_axes(knl, "res", "N2,N0,N1")
\end{lstlisting}
\end{tcolorbox}
\noindent
These transformations change the nesting order of the data axes, \texttt{N0}
and \texttt{N1}, which changes the global memory access patterns for
\texttt{u} and \texttt{res} so that the stride of \texttt{lid(0)} is 1
instead of \texttt{nunit\_nodes}. This significantly improves the performance
of these loads. As a result, the last variant is the fastest in all our
measurements, and achieves between 5\% and 18\% of peak FLOP/s rates on all
five GPUs.

\begin{figure*}
\centering
\begin{subfigure}{.345\textwidth}
    \includegraphics[trim={.25cm 0 .25cm .25cm},clip,width=\textwidth,right]{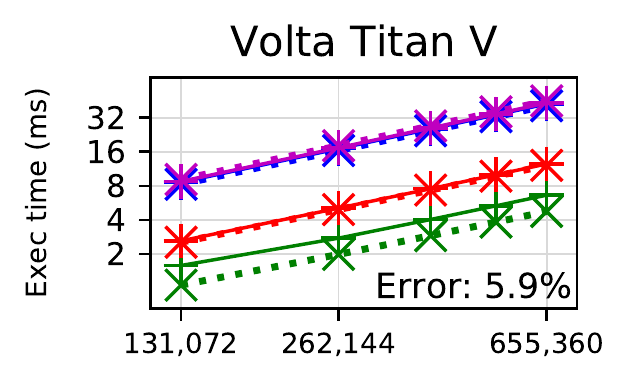}
\end{subfigure}
\begin{subfigure}{.315\textwidth}
    \includegraphics[trim={.25cm 0 .25cm .25cm},clip,width=\textwidth,right]{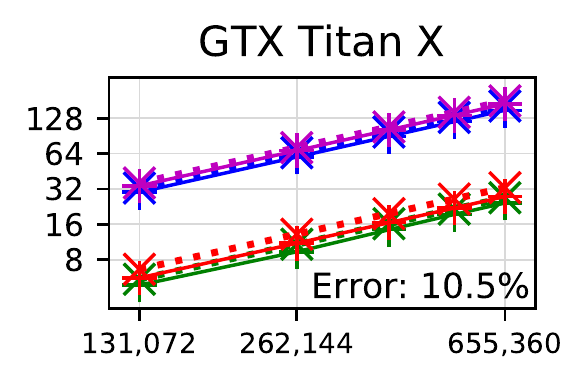}
\end{subfigure}
\begin{subfigure}{.315\textwidth}
    \includegraphics[trim={.25cm 0 .25cm .25cm},clip,width=\textwidth,right]{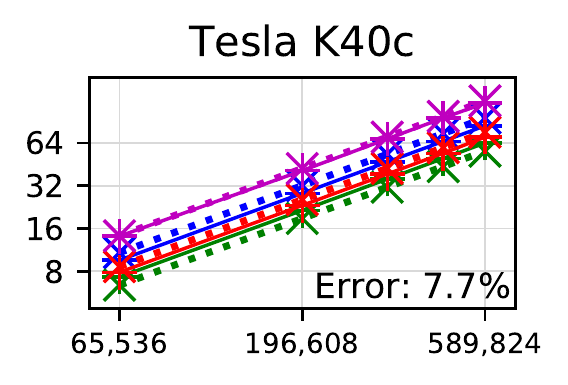}
\end{subfigure}
\begin{subfigure}{.345\textwidth}
    \includegraphics[trim={.25cm 0 .25cm .25cm},clip,width=\textwidth,right]{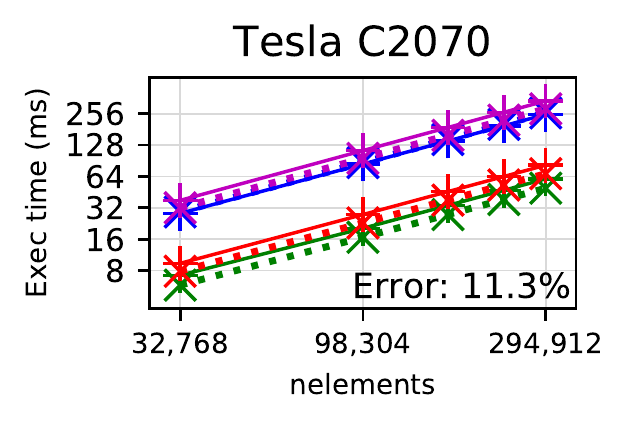}
\end{subfigure}
\begin{subfigure}{.315\textwidth}
    \includegraphics[trim={.25cm 0 .25cm .25cm},clip,width=\textwidth,right]{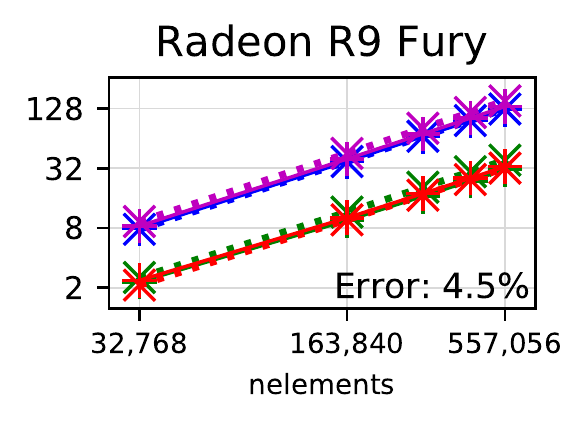}
\end{subfigure}
\begin{subfigure}{.315\textwidth}
    \centering
    \vspace{0.15cm}
    \setlength{\tabcolsep}{0.08cm}
\resizebox{0.96\linewidth}{!}{
\begin{tabular}{lrrrrr}
  {\small Variant} & {\small TiV} & {\small TiX} & {\small K40} & {\small 2070} & {\small Fury} \\
  \midrule
  {\color{plotblue} \small PFdm} & 3.6 & 8.6 & 13.3 & 3.3 & 0.4 \\
  {\color{plotgreen} \small PFdm\textsuperscript{T}} & 29.1 & 12.4 & 12.3 & 18.5 & 19.8 \\
  {\color{plotred} \small PFu} & 2.8\rlap{\textsuperscript{L}} & 19.1 & 9.4\rlap{\textsuperscript{L}} & 16.9\rlap{\textsuperscript{L}} & 4.5 \\
  {\color{plotmagenta} \small NoPF} & 4.2 & 6.0 & 2.3 & 15.8 & 13.2 \\
  \addlinespace[1pt]
  {\small Mean} & \textbf{5.9} & \textbf{10.5} & \textbf{7.7} & \textbf{11.3} & \textbf{4.5} \\
\end{tabular}
}
\setlength{\tabcolsep}{\defaulttabcolsep}
 \\
    \vspace{0.15cm}
    \begin{tikzpicture}
      [
        nodestyle/.style={font=\footnotesize,thick,inner sep=1pt}
      ]

        \node [nodestyle, anchor=west] (Noprefetch) at (0.5, 0) {No prefetch};
        \draw [draw, thick, plotmagenta] (0, 0) -- (Noprefetch.west);
        \node [nodestyle, anchor=west] (Prefetchdiffmat) at (2.8, 0) {Prefetch diff mat};
        \draw [draw, thick, plotblue] (2.3, 0) -- (Prefetchdiffmat.west);
        \node [nodestyle, anchor=west] (Prefetchu) at (.5, -.4) {Prefetch u};
        \draw [draw, thick, plotred] (0, -.4) -- (Prefetchu.west);
        \node [nodestyle, anchor=west] (PrefetchdiffmatT) at (2.8, -.4) {Prefetch diff mat\textsuperscript{T}};
        \draw [draw, thick, plotgreen] (2.3, -.4) -- (PrefetchdiffmatT.west);

        \node [nodestyle, anchor=west] (Measured) at (0.5, -.8) {Measured};
        \draw [draw, thick] (0,-.8) -- (Measured.west);
        \node [nodestyle, anchor=west] (Predicted) at (2.8, -.8) {Predicted};
        \draw [draw, thick, dashed] (2.3,-.8) -- (Predicted.west);
        \draw (0.25,-.8) node[cross=0.12cm, rotate=45] {};
        \draw (2.55,-.8) node[cross=0.14cm] {};
    \end{tikzpicture}
\end{subfigure}
\caption{DG differentiation model accuracy. The table displays the geometric
    mean of relative error (\%). \textsuperscript{T}Element data transposed.
\textsuperscript{L}Linear model used (otherwise nonlinear).}
\label{fig:dg_results}
\end{figure*}
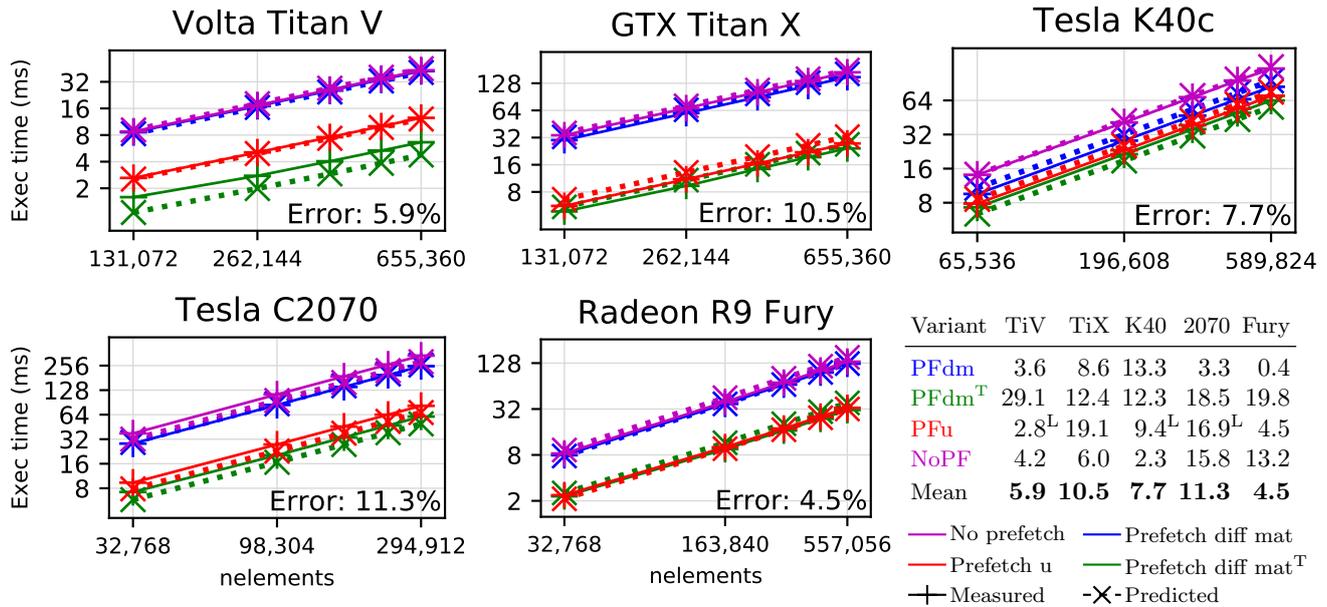

All four variants operate on 32-bit floating point data and use $16\times16$
work-groups. We set \texttt{nmatrices} to $3$, \texttt{nunit\_nodes} to $64$,
and \texttt{nelements} varies as shown in Figure~\ref{fig:dg_results}. The four
variants and the measurement kernel set used to calibrate their model use 11 distinct
global memory access patterns as shown in Figure~\ref{fig:mkernels_other}.

To decide whether to use a model that allows for on-chip cost hiding, we apply
the analysis described in Section~\ref{sec:results_models} to each of the DG
variants. The results suggest that on-chip work overlaps with global memory
transactions, with one exception: the \texttt{u}-prefetching variant does not
exhibit this overlap on the \nvidiavolta, \nvidiakepler, and \nvidiafermi\ GPUs.
On these three GPUs, the total execution time for the \texttt{u}-prefetching
variant is approximately the sum of the on-chip and global memory operation
costs. Because of this, we use the linear model expressed in
\eqref{eq:linear_model} to model the \texttt{u}-prefetching variant on
these three GPUs, and in all other cases, we model the DG variants using the
\emph{nonlinear} model expressed in \eqref{eq:nonlinear_model}. The
model features comprising $c_\text{gmem}$, $c_\text{on-chip}$, and
$c_\text{overhead}$, as well as the measurement kernel set, are shown in
Figure~\ref{fig:mkernels}. Recall that in Section~\ref{sec:modeling_overlap}
we observed that the kernel which allowed us to vary the ratio of local to
global memory accesses also did not exhibit overlap on the \nvidiakepler\ and
\nvidiafermi\ GPUs.

Figure~\ref{fig:dg_results} compares modeled to measured execution times for the
four variants on five GPUs, and displays the relative error in model predictions.
Across all cases the geometric mean of relative error is 7.5\%. On all four
Nvidia GPUs the model predictions are sufficient to accurately rank execution
times for all variants from highest to lowest. On the \amdradeon\ GPU, while the
model predictions would rank the two fastest variants in reverse order, the
predicted execution times are accurate enough to narrow the space of potential
variants to the two fastest options, whose execution times differ by less than
7\%. Additionally, the predictions accurately reveal the cost savings realized
by the \texttt{diff\_mat}-prefetching variant when operating on element data
with a transposed memory layout.

\subsection{Finite Differences}
\label{sec:fd_results}

In our third evaluation case, we model execution times for two variants of a
2-D five-point finite difference stencil operation. The pre-transform \loopy\
kernel shows the mathematical operation carried out by both transform variants:
\begin{tcolorbox}[listingbox]
\begin{lstlisting}[style=custompython]
knl = lp.make_kernel(
   "{[i,j]: 0<=i,j<n}",
   "res[i,j] = u[i,j+1] + u[i+1,j] - 4*u[i+1,j+1] + u[i+1,j+2] + u[i+2,j+1]")
\end{lstlisting}
\end{tcolorbox}

Both variants parallelize the \texttt{i} and \texttt{j} indices across threads
and prefetch \texttt{lsize[0]} $\times$ \texttt{lsize[1]} tiles from \texttt{u}
into scratchpad (local) memory before performing arithmetic:
\begin{tcolorbox}[listingbox]
\begin{lstlisting}[style=custompython]
knl = lp.split_iname(knl,
    "i", lsize[1]-2, outer_tag="g.1", inner_tag="l.1")
knl = lp.split_iname(knl,
    "j", lsize[0]-2, outer_tag="g.0", inner_tag="l.0")
knl = lp.add_prefetch(knl,
    "u", ["i_in", "j_in"], fetch_bounding_box=True)
knl = lp.tag_inames(knl, "u_dim_0:l.1, u_dim_1:l.0")
\end{lstlisting}
\end{tcolorbox}

The difference between
the two variants is the work-group size, which is also the size of the tiles
fetched into local memory. This change substantially affects the global
memory access patterns. For the first variant, we use $16\times16$
work-groups, and for the second, we use $18\times18$. With $16\times16$
work-groups, $16\times16$-element tiles are prefetched, with one fetch per
thread. After this fetch, the result for each of the interior $14\times14$
elements is computed by one of 196 threads, while 60 threads remain idle,
corresponding to the 60 halo elements. This strategy yields a \texttt{gid(0)}
stride of 14. With $18\times18$ work-groups, $18\times18$-element tiles are
prefetched, with one fetch per thread, and the result for each of the interior
$16\times16$ elements is computed by one of 256 threads, while 68 threads
remain idle, corresponding to the 68 halo elements. This strategy yields a
\texttt{gid(0)} stride of 16.

Unlike the other variants we model, the global memory loads in these kernels
have access-to-footprint ratios near one. Because of this, the data throughput,
calculated as $(\text{total global memory access count})/(\text{execution
time})$, is less likely to be inflated significantly by cached data reuse, and
is meaningful to report. We observe that the $16\times16$ variant is slightly
faster, and achieves between 40\% and 82\% of peak bandwidth on all five GPUs.
The effective FLOP/s rates are between 2\% and 5\% of peak. 

The analysis of on-chip cost hiding that we described in
Section~\ref{sec:results_models} indicates that little if any such overlap
occurs when executing these variants on the architectures used in our
experiments. Because of this, we model execution time using the
\emph{linear} model in \eqref{eq:linear_model}. The model
features comprising $c_\text{gmem}$, $c_\text{on-chip}$, and
$c_\text{overhead}$, as well as the measurement kernel set, are shown in
Figure~\ref{fig:mkernels}. As shown in Figure~\ref{fig:mkernels_other}, the
variants and the measurement kernels used to calibrate the model are based on
five distinct global memory access patterns. Both variants operate on
32-bit floating point data.



We note two potential sources of modeling error in this example. As mentioned in
Section~\ref{sec:mkernel_design}, the models presented here use a single feature
to represent all local memory accesses. We made this decision to simplify
the models and measurement kernel sets; it is not a limitation of
our approach or our software, since local memory access features may include the
same access pattern characteristics as global memory access. The local memory
accesses in these kernels constitute a significant portion of the execution
time, 10-20\%, and have different access patterns across variants. The
\texttt{lid(0)} stride is one in both variants; however, the \texttt{lid(1)}
stride is 18 in the $18\times18$-tile variant, and 16 in the $16\times16$-tile
variant. If these accesses differ in cost from one another, or from those in
the local memory access measurement kernel, this model cannot account for these
differences.

Another potential source of error for this example is the effect of varying
machine utilization on execution time, which our approach does not attempt to
capture, as discussed in Section~\ref{sec:assumptions}. The amount of local
memory used per work-group, as well as the number of threads per work-group,
both differ between these two variants, and can affect machine utilization,
e.g., through varying amounts of latency hiding.

\begin{figure*}
\centering
\begin{subfigure}{.345\textwidth}
    \includegraphics[trim={.25cm 0 0 .25cm},clip,right]{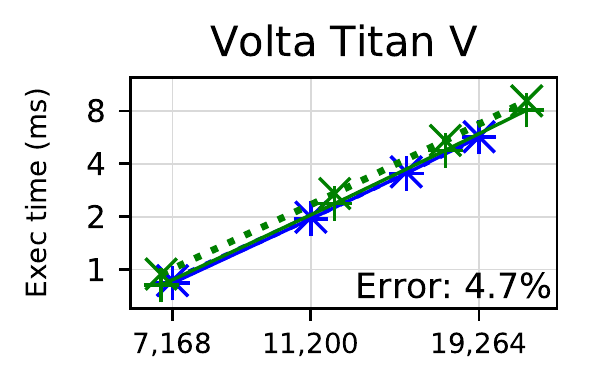}
\end{subfigure}
\begin{subfigure}{.315\textwidth}
    \includegraphics[trim={.25cm 0 0 .25cm},clip,right]{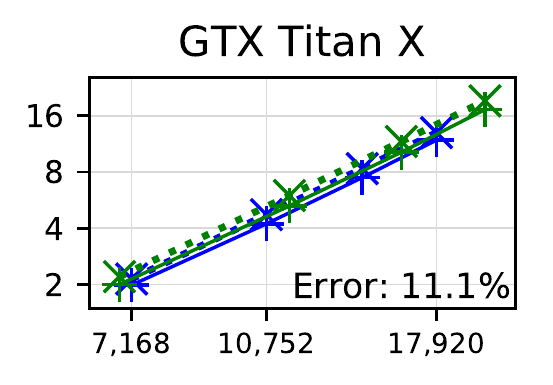}
\end{subfigure}
\begin{subfigure}{.315\textwidth}
    \includegraphics[trim={.25cm 0 0 .25cm},clip,right]{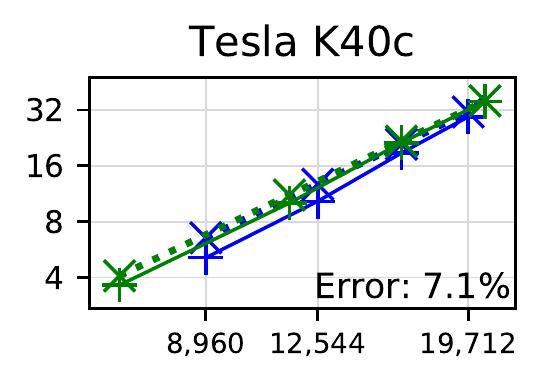}
\end{subfigure}
\begin{subfigure}{.345\textwidth}
    \includegraphics[trim={.25cm 0 0 .25cm},clip,right]{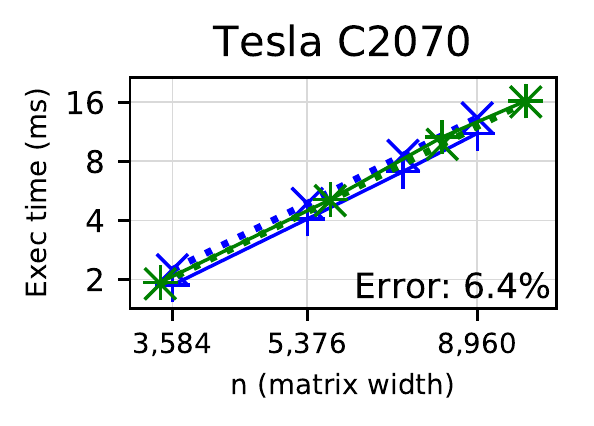}
\end{subfigure}
\begin{subfigure}{.315\textwidth}
    \includegraphics[trim={.25cm 0 0 .25cm},clip,right]{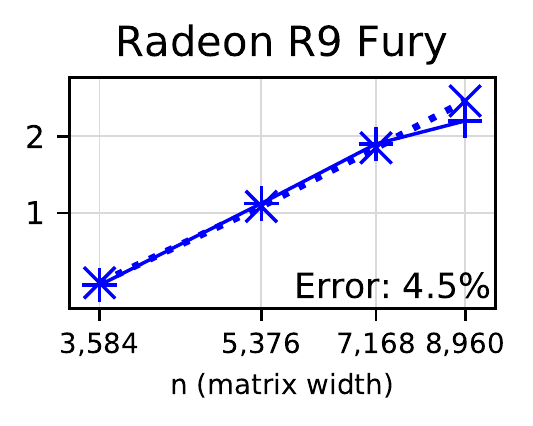}
\end{subfigure}
\begin{subfigure}{.315\textwidth}
    \centering
    \setlength{\tabcolsep}{0.08cm}
\resizebox{0.96\linewidth}{!}{
\begin{tabular}{lrrrrr}
  {\small Variant} & {\small TiV} & {\small TiX} & {\small K40} & {\small 2070} & {\small Fury} \\
  \midrule
  {\color{plotblue} \small 16x16} & 1.6 & 10.1 & 14.8 & 20.2 & 4.5 \\
  {\color{plotgreen} \small 18x18} & 14.0 & 12.2 & 3.4 & 2.1 &  \\
  \addlinespace[1pt]
  {\small Mean} & \textbf{4.7} & \textbf{11.1} & \textbf{7.1} & \textbf{6.4} & \textbf{4.5} \\
\end{tabular}
}
\setlength{\tabcolsep}{\defaulttabcolsep}
 \\
    \vspace{0.2cm}
    \begin{tikzpicture}
      [
        nodestyle/.style={font=\footnotesize,thick,inner sep=1pt}
      ]

        \node [nodestyle, anchor=west] (16x16) at (0.5, 0) {16x16};
        \draw [draw, thick, plotblue] (0, 0) -- (16x16.west);
        \node [nodestyle, anchor=west] (18x18) at (2.8, 0) {18x18};
        \draw [draw, thick, plotgreen] (2.3, 0) -- (18x18.west);

        \node [nodestyle, anchor=west] (Measured) at (0.5, -.4) {Measured};
        \draw [draw, thick] (0,-.4) -- (Measured.west);
        \node [nodestyle, anchor=west] (Predicted) at (2.8, -.4) {Predicted};
        \draw [draw, thick, dashed] (2.3,-.4) -- (Predicted.west);
        \draw (0.25,-.4) node[cross=0.12cm, rotate=45] {};
        \draw (2.55,-.4) node[cross=0.14cm] {};
    \end{tikzpicture}
\end{subfigure}
\caption{Finite difference model accuracy. The table displays the geometric mean of relative error (\%).}
\label{fig:fd_results}
\end{figure*}
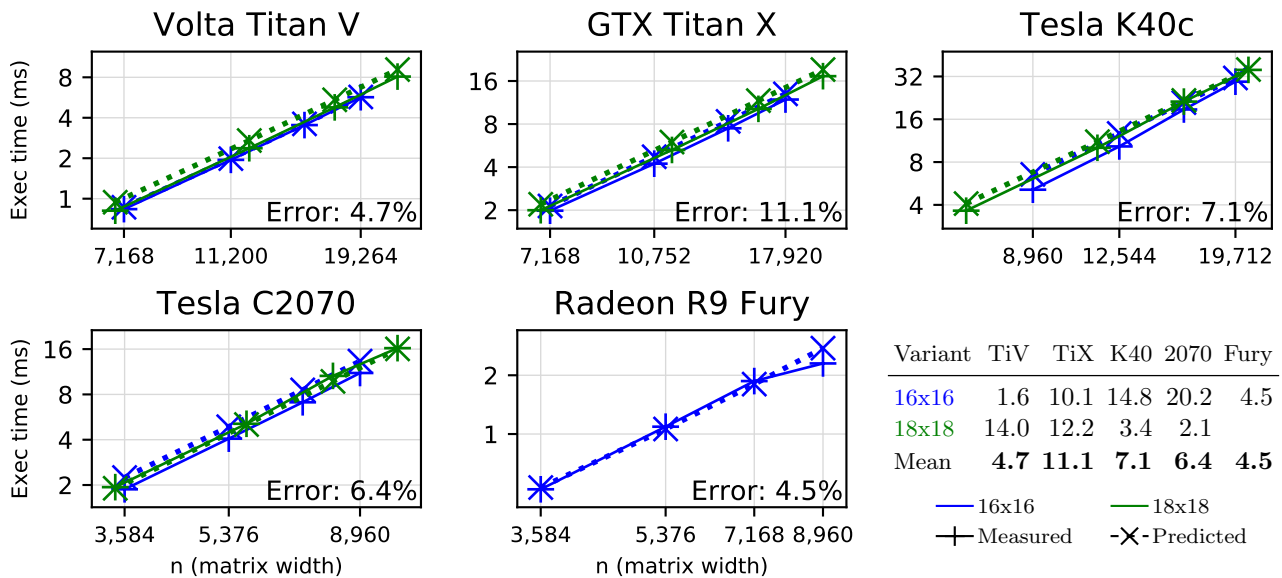

Figure~\ref{fig:fd_results} compares modeled to measured execution times for the
two variants on five GPUs, and displays the relative error in model predictions.
Note that the 256 work-item limit on the AMD GPU prevents us from executing the
$18\times18$-tile variant. Despite the potential sources of error described
above and the similarity in execution times between the two variants, across all
cases the geometric mean of relative error is 6.7\%, and the model predictions
are sufficiently accurate to identify the faster variant in every case except
for the \nvidiafermi\ GPU. 

\section{Related Work}
\label{sec:related_work}

Common approaches to cost prediction include analytical modeling (often
based on in-depth program and hardware analysis), statistical regression and
machine learning approaches, and machine and application benchmarking. Many
approaches use a combination of these strategies.

Among analytical approaches to GPU performance modeling, much of the previous
work yielding the most accurate predictions has focused on constructing
models of instruction-level execution based on detailed hardware knowledge
and instruction analysis, often for a single architecture or group of highly
similar architectures. Many of these models predict well for their specific
target architecture. For example, \cite{hong_analytical_2009} present an
analytical performance model for Nvidia GPU architectures that produces an
execution time prediction based on estimates of memory-level and thread-level
parallelism. They further extend their model for power prediction
\citep{hong_integrated_2010}. This model achieves a geometric mean error of
13.3\% when predicting performance of the Merge \citep{linderman_merge:_2008}
benchmarks on four Tesla generation Nvidia GPUs. It makes extensive use of
hardware performance characteristics, such as timing delays between memory
transactions, DRAM (dynamic random-access memory) access latency, and
instruction execution cycles, and it requires an analysis of PTX (Parallel
Thread Execution) assembly instructions. \cite{baghsorkhi_adaptive_2010} also
use deep analytical knowledge of a (single) GPU architecture, and, unlike
\cite{hong_analytical_2009}, model branch divergence, bank conflicts, and SIMD
pipeline delays.

\cite{spafford2012aspen} approach analytical modeling of exascale
applications by introducing a domain specific language for defining 
abstract application and machine models. To construct an application model,
users provide a structured description of application characteristics,
including parameterized expressions for FLOP counts, load counts,
communication volume and type (e.g., all-to-all vs. allgather), and the
number of parallel units, as well as information on control-flow. To
construct an abstract machine model, users provide a structured description
of machine characteristics, including memory clock rate, bus width, core
clock rate, SIMD width, and presence of an FMA instruction,
as well as information on the machine interconnect. These abstract models can
then be consumed by other applications to analyze the code and its
performance.

Other related analytical modeling contributions include works by
\cite{hammer2017kerncraft}, who use a partially automated analytical approach
to modeling CPU loop kernel performance that allows for multiple architectures,
\cite{gemund2003symbolic}, who uses a partially automated symbolic analytic
modeling approach to predict performance on distributed CPU machines,
\cite{pllana2005performance}, who provide a graphical user interface to aid
distributed CPU model creation and employ discrete event simulation, and
\cite{unat2015exasat}, who introduce a tool employing compiler analysis to
generate parameterized models with the slightly different goal of
evaluating design trade-offs and software optimizations. All four of these
analytical approaches require a user-supplied machine model or architecture
statistics.

Machine learning and statistical techniques are also used to predict
performance of GPU kernels. From the perspective of optimization selection,
\cite{cavazos_automatic_2006} present a probabilistic predictor of
transformation selection using a non-analytical, black-box model based on an
artificial neural network. \cite{joseph_predictive_2006} use techniques from
machine learning to identify piecewise nonlinearities in cost metrics. Other
approaches emphasize the performance of single subsystems, such as branch
prediction \citep{emer_asim_2002}. Other learning and statistical approaches
include \cite{jia2012stargazer}, \cite{kerr2012eiger}, \cite{wu2015ml},
\cite{zhang2011perfpower}, \cite{chen_learning_2018}, and
\cite{gysi2019absinthe}. 


Some modeling approaches employ benchmarks, including
\cite{zhang_quantitative_2011}, who use results from microbenchmarks to derive a
throughput model for instruction pipeline, shared memory, and global memory
costs. They focus on identifying performance bottlenecks and guiding the
optimization process rather than predicting execution time. The target kernel
must be run in a simulator to gather relevant performance counters, and the
binary file must be analyzed. 
\cite{johnston2018opencl} gather their set of architecture-independent program
features by simulating an OpenCL device using the Oclgrind simulator and
examining the LLVM intermediate representation produced. They then use these
features to generate a random forest model. 
\cite{konstantinidis2017aquantitative} employ a microbenchmarking approach to
gather GPU performance metrics, and gather information about their target
kernels using Nvidia's \texttt{nvprof} profiler on a reference GPU. They
show results on a larger kernel set than most works we found, achieving
an average error of 27.66\% (geometric mean 18\%, not reported).


Two recent articles survey the current GPU performance modeling landscape. In
\cite{madougou2016landscape}, researchers perform an in-depth evaluation of 12
GPGPU performance modeling tools, including 6 analytical models:
\cite{hong_analytical_2009}, \cite{kothapalli2009model}, \cite{li2015visual},
\cite{meswani2013modeling}, \cite{sim2012framework}, and
\cite{song2013simplified}. They determine that, while the analytical models
tend to be accurate for a particular hardware family or workload, they are less
accurate for different hardware. They report that constructing all of these
models requires significant effort and the collection or estimation of anywhere
from 6 to 30 parameters characterizing the hardware or the application, which
is consistent with our experience reproducing the results in
\cite{hong_analytical_2009}.


\cite{lopez-novoa2015survey} come to similar conclusions after surveying over 30
models for GPU computing of various types, including 7 general-purpose
execution-time-estimating models, i.e., models that are not designed for a
particular application
\citep{hong_analytical_2009,kothapalli2009model,kerr2010modeling,che2014benchfriend,garcia2013modelo,meng2011grophecy,nugteren2012boat}.
They conclude that there is no accurate model valid for a wide set of
architectures, and that each model they consider makes a trade-off between
accuracy and breadth of hardware applicability. They also note that most models
they surveyed are designed for CUDA rather than vendor-neutral OpenCL, and that
\cite{hong_analytical_2009} stands out as the model of choice for accurately
predicting GPU execution times.





\section{Conclusions}
\label{sec:conclusions}

We have demonstrated an alternative to previous GPU performance modeling
approaches: a framework for constructing analytical models and calibrating them
to a GPU using a customized measurement kernel set. Our framework allows a
developer to control the trade-offs between model accuracy, complexity,
generalization, and evaluation speed, and our hardware-blind model-calibration
approach allows these models to make predictions on new devices with minimal
effort. We demonstrate example execution time models for three workloads
yielding predictions accurate enough to, e.g., allow an autotuner or human user
to identify which kernel variant or subset of variants will have the shortest
execution times. Across all variants of all three computations on all five
GPUs, we achieved a geometric mean relative error of 6.4\%. Additionally, we
show how the transparency and interpretability of the model expressions,
parameters, and features enables users to gain actionable insights into the
factors affecting computational cost.


\begin{acks}

The authors would like to thank Dr. Bill Gropp for helpful discussions and
feedback on an earlier version of the manuscript, and express appreciation to
Matt Wala for significant contributions to our underlying transformation engine
\loopy\ and unfailing readiness to assist.

The author's work was supported in part by US Navy ONR grant numbers
N00014-14-1-0117, and by the National Science Foundation under grant numbers
DMS-1418961 and CCF-1524433. We also gratefully acknowledge a hardware gift
from Nvidia Corporation.

Opinions expressed herein are those of the author and in no way reflect the
official position of any of the funding agencies.

\end{acks}
\FloatBarrier
\bibliographystyle{SageH}
\bibliography{refs}

\end{document}